\documentstyle[aps,prc,epsfig,floats,preprint]{revtex}

\begin{document}
\draft
\tightenlines

\preprint{NT@UW-99-63}
\title{Ground states of the Wick-Cutkosky model using light-front
dynamics}
\author{Jason R.~Cooke and Gerald A.~Miller}
\address{
Department of Physics \\
University of Washington \\
Box 351560 \\
Seattle WA 98195-1560, USA}
\date{\today{}}
\maketitle

\begin{abstract}
We consider the ground state in a model with scalar nucleons and a meson
using the formalism of light-front dynamics.
Light-front potentials for two-nucleon bound states are calculated
using two approaches.
First, light-front time-ordered perturbation theory is used to calculate
one- and two-meson-exchange potentials.  These potentials give results
that agree well with the ladder and ladder plus crossed box
Bethe-Salpeter spectra.
Secondly, approximations that incorporate non-perturbative physics are
used to calculate alternative one-meson-exchange potentials.  These
non-perturbative potentials give better agreement with the spectra of
the full non-perturbative ground-state calculation than the perturbative
potentials.  For lightly-bound states, all of the approaches appear to
agree with each other.
\end{abstract}

\pacs{PACS number(s):
21.45.+v, 
03.65.Ge, 
03.65.Pm, 
11.10.Ef  
}

\section{Introduction}

Recent experiments at Thomas Jefferson National Accelerator Facility
have measured the $A(Q^2)$ structure function of the deuteron for
momentum transfers up to 6 (GeV/c)$^2$ \cite{Alexa:1999fe}, and
measurements for $B(Q^2)$ are planned.  At such large momentum
transfers, a relativistic description of the deuteron is required.  One
approach that gives such a description is light-front dynamics, which we
will examine here.  To separate the effects of the using light-front
dynamics from the effects of the model, we choose to use the massive
Wick-Cutkosky model.  This is a ``toy model'' investigation, instead of
the full nuclear theory calculation.  Using this model, the light-front
Hamiltonian approach is used to solve for the bound-state wavefunction.
The results of our calculation can then be compared to other
calculations done with the same model but different approaches.  The
simplest observable that can be compared is the relationship between the
bound-state mass and the coupling constant.

The utility of the light-front dynamics was first discussed by Dirac
\cite{Dirac:1949cp}.  We start by expressing the four-vector $x^\mu$ in
terms of the light-front variables $x^\mu=(x^+,x^-,x^1,x^2)$, where
$x^\pm=x^0\pm x^3$.  This is simply a change of variables, but an
especially convenient one.  Using this coordinate system and defining the
commutation relations at equal light-front time ($x^+=t_{\text{LF}}$),
we obtain a light-front Hamiltonian
\cite{Brodsky:1998de,Harindranath:1996hq,Heinzl:1998kz}.  The
Hamiltonian is used in the light-front Schr\"odinger equation to solve
for the ground state.

There are many desirable features of the light-front dynamics and the
use of light-front coordinates.  
First of all, high-energy experiments are naturally described
using light-front coordinates.  The wave front of a beam of high-energy
particles traveling in the (negative) three-direction is defined by a
surface where $x^+$ is (approximately) constant.  Such a beam can probe
the wavefunction of a target described in terms of light-front variables
\cite{Brodsky:1998de,Miller:1997cr}: the Bjorken $x$ variable used to
describe high-energy experiments is simply the ratio of the plus
momentum of the struck constituent particle to the total plus momentum
($p^+$) of the bound state.
Secondly, the vacuum for a theory with massive particles 
can be very simple on the light front.  This is
because all massive particles and anti-particles have positive plus
momentum, and the total plus momentum is a conserved quantity.  Thus,
the na\"{\i}ve vacuum (with $p^+=0$) is empty, and diagrams that couple
to this vacuum are zero.  This greatly reduces the number of non-trivial
light-front time-ordered diagrams.
Thirdly, the generators of boosts in the one, two, and plus directions
are kinematical, meaning they are  independent of the interaction.
Thus, even when the Hamiltonian is truncated, the wavefunctions will
transform correctly under boosts.  Thus, light-front dynamics is useful
for describing form factors at high momentum transfers.
A drawback of the light-front formalism is that the Hamiltonian is not
manifestly rotationally invariant, since the generators of rotations
about the one and two directions are dynamical.  A study of the effects
of the loss of rotational invariance of the excited states in the model
being used here was made in Ref.~\cite{Cooke:1999yi}, which shows that
there is less breaking of the degeneracy in the spectrum when higher
order potentials are used as opposed to lower order potentials.

There are other approaches that can be used to obtain relativistic
wavefunctions and bound-state energies, including the Feynman-Schwinger
representation (FSR) of the two-particle Green's
function\cite{Simonov:1993kp} and the Bethe-Salpeter equation (BSE)
\cite{Nambu:1950vt,Schwinger:1951ex,Gell-Mann:1951rw,Salpeter:1951sz}.
The FSR is useful since it can be constructed so it is equivalent to the
Bethe-Salpeter equation using a kernel where {\em all}
two-particle-to-two-particle ladder and crossed ladder diagrams are
included.  However, it requires a path integral to be done numerically,
so it is computationally intensive to obtain an accurate answer.  The
FSR result is to be considered the full solution that the Bethe-Salpeter
and Hamiltonian equations approximate.
The BSE can be solved much quicker than the FSR, however a truncation of
the BSE kernel is required, causing the BSE results to differ from the
FSR results.  It is well known that any finite truncation of the kernel
yields bound-state wavefunctions with problems, such as the incorrect
one-body limit \cite{Gross:1982nz}.  For our scalar model
the potential is always attractive, so truncation of the kernel gives
binding energies that are too small
\cite{LevineWright1,LevineWright2,LevineWright3}.  
Another approach is explicitly covariant light-front dynamics
\cite{Carbonell:1998rj,Schoonderwoerd:1998pk}, where manifest covariance
is kept at the price of using a null-plane whose orientation is not
fixed.

Here we study light-front dynamics because of its close connection to
experimental observables.  Using field theory, light-front potentials
can be derived that give results physically equivalent to those of the
Bethe-Salpeter equation.  We define ``physically equivalent'' in section
\ref{tbe:sbSec}.
Depending on the diagrams used to construct the potential, one can argue
that certain potentials are physically equivalent to the Bethe-Salpeter
equations with certain kernels.  The best that these potentials can do
is reproduce the results of the corresponding Bethe-Salpeter equations.
On the other hand, light-front potentials can be constructed that
attempt to incorporate non-perturbative physics.  It is possible that
these potentials can give results that agree with the full theory
results better than the Bethe-Salpeter equation using low-order
kernels.  We will consider both types of potentials in this paper, and
see how well they perform.

A brief discussion about the approximation used is in order.
It is well known that the vacuum of the full Wick-Cutkosky model is
unstable \cite{Baym} due to the cubic coupling which provides the
interaction.  However, when the bound-state calculation is restricted to
the two-particle sector, the quenched approximation is used, and the
self-energy and vertex-renormalization diagrams are neglected, the
theory has a well defined ground state.  In this paper, we compare the
results of our light-front Hamiltonian calculation to the Bethe-Salpeter
and FSR calculations, both of which use the same
approximations.  The use of these simplifying approximations allows us to
highlight the differences between the various approaches.
The inclusion of the self-energy diagrams and counterterms for the
light-front Hamiltonian \cite{Wivoda:1993qr} and for the FSR
\cite{Savkli:1999rw,Rosenfelder:1996bd} will not be discussed here.

\subsection{Outline of the paper}

The objective of this paper is to obtain the bound-state energy for the
ground state in our theory.  In our model, neutral scalar nucleons
interact via a Yukawa interaction, which is mediated by a neutral scalar
meson.  The light-front Hamiltonian derived from the Lagrangian is used
in the light-front Schr\"odinger equation for a two-nucleon bound state.
The rules for light-front time-ordered perturbation theory
(LFTOPT) are then derived for this Schr\"odinger equation, along with
the Feynman rules for the effective potential.  All of this is discussed
in more detail in section \ref{ourmodel}.

The full potential for the light-front Schr\"odinger equation is given
by an infinite sum of diagrams.  Using the LFTOPT rules in section
\ref{apptrunc}, we derive the one-boson-exchange (OBE) and
two-boson-exchange (TBE) potentials, where the diagrams that give rise
to mass and vertex renormalization are not included.
The two-boson-exchange stretched-box (TBE:SB) diagrams, a subset of the
TBE diagrams, are used to construct the TBE:SB potential, and the
utility of this potential is commented on.  After the discussion of the
perturbative potentials, we define several potentials in section
\ref{appphys} which attempt to incorporate non-perturbative physics in a
OBE type of potential.  Three potentials are obtained by approximating
the OBE potential directly (giving the symmetrized mass, instantaneous,
and retarded potentials), and one potential is obtained by approximating the
Bethe-Salpeter equation and reducing it to a three-dimensional equation
(giving the modified-Green's-function potential).

In section \ref{results}, the potentials are used to numerically obtain
the spectra, the coupling constant versus bound-state mass curves.
This is done by solving the light-front Schr\"odinger equation with each
of the truncated potentials (OBE, TBE, and TBE:SB), and the approximate
potentials (symmetrized mass, instantaneous, retarded, and
modified-Green's-function).  We compare these results to those in the
literature obtained with other approaches.

We summarize our findings in section \ref{conclusions}.  Only a few
low-order terms from the perturbative potentials are needed to
approximate the results of the physically equivalent Bethe-Salpeter
equations \cite{Cooke:1999yi,Sales:1999ec,Ligterink:1995tm}. Since the
light-front potentials are calculated without including the mass and
vertex renormalization diagrams, the Bethe-Salpeter kernels used also do
not include those diagrams.  Since the interaction in the Wick-Cutkosky
model is strictly attractive, the spectra calculated using the perturbative
potentials will underestimate the the binding energy compared with the 
spectra for the physically equivalent Bethe-Salpeter equations.  As
progressively higher-order terms in the potential are kept, the spectra
calculated will agree better with the BSE spectra.
However, when a truncated kernel is used in the Bethe-Salpeter equation,
the solutions obtained are known to be a poor approximation of the full
solution \cite{LevineWright1,LevineWright2,LevineWright3}, which for our
theory (where no renormalization graphs are kept) is given by the
Feynman-Schwinger representation of the Green's 
function \cite{Simonov:1993kp,Nieuwenhuis:1996mc}.  Hence, our truncated
potentials cannot give the full results.  The non-perturbative
approximations of the potential give spectra that more closely match the
spectrum of the full solution.  This suggests that these approximations
are reproducing the physics more accurately than the perturbative
potentials.

The conventions and notations employed in this paper are
summarized in Appendix \ref{notation}.  The conversion of the
light-front Schr\"odinger equation into matrix form suitable for 
numerical evaluation is discussed in Appendix \ref{numericaleq}.
Azimuthal-angle
integrations of the OBE potentials, which help simplify the evaluation
of the bound-state wavefunctions, are given in Appendix \ref{angint}.
The loop integrations and azimuthal-angle integrations needed for the TBE
potentials are discussed in Appendix \ref{loopints}.  A check of the
validity of the uncrossed approximation used in section \ref{sect:3dred}
is done in Appendix \ref{uxcheck}.

This study is related to the work of Sales {\it et al.}
\cite{Sales:1999ec}, who computed bound states using the light-front
Hamiltonian with the OBE and TBE:SB approximations, and compared to the
ladder Bethe-Salpeter equation results.  Here, we consider
also the effect of including the crossed graph part of the TBE potential
as well as several non-perturbative approximations. 

\section{Our model} \label{ourmodel}

We consider an isospin doublet of two uncharged scalars
$\phi=(\phi_1,\phi_2)$ with mass $M$ (which we will refer to as {\em
nucleons}), that couple to a third, uncharged scalar $\chi$ with mass
$\mu$ (which we will refer to as a {\em meson}) by a $\phi^2 \chi$
interaction.  This is the massive extension of the Wick-Cutkosky model
\cite{WickCutkosky}, which has been used on the light front to study
scattering states \cite{Schoonderwoerd:1998pk} as well as bound states
\cite{Sales:1999ec}.  The Lagrangian is
\begin{eqnarray}
{\mathcal L} &=& 
\frac{1}{2} \left( \partial_\mu \phi \partial^\mu \phi -   M^2 \phi^2 \right) +
\frac{1}{2} \left( \partial_\mu \chi \partial^\mu \chi - \mu^2 \chi^2 \right) +
g\frac{M}{2} \phi^2 \chi, \label{thelagrangian}
\end{eqnarray}
where $g$ is a dimensionless coupling constant and
$\phi^2=\phi_1^2+\phi_2^2$.

\subsection{Light-front Hamiltonian}

To obtain the light-front Hamiltonian from the Lagrangian in
Eq.~(\ref{thelagrangian}), we follow the approach used by Miller
\cite{Miller:1997cr} and many others (see the review
\cite{Brodsky:1998de}) to write the light-front Hamiltonian ($P^-$) as
the sum of a free, non-interacting part and a term containing the
interactions.  We use the conventions given in Appendix
\ref{notation}. The operators we use can be expressed in terms of Fock
space operators since for this theory in light-front dynamics, the
physical vacuum is the Fock space vacuum, and thus the Hilbert space is
simply the Fock space.  The Hamiltonian is obtained by using the
energy-momentum tensor in
\begin{eqnarray}
P^\mu=\frac{1}{2}\int dx^-d^2x_\perp\;
T^{+\mu}(x^+=0,x^-,\bbox{x}_\perp),
\end {eqnarray}
The usual relations determine  $T^{+\mu}$, with
\begin{eqnarray}
T^{\mu\nu}=-g^{\mu\nu}{\cal L} +\sum_r{\partial{\cal L}\over\partial
  (\partial_\mu\phi_r)}\partial^\nu\phi_r,
\label{tmunu}
\end{eqnarray}
in which the degrees of freedom (the fields $\phi$ and $\chi$) are
labeled by $\phi_r$.

It is worthwhile to consider the limit in which the interactions between the 
fields are removed.  This will allow us to define the free Hamiltonian $P^-_0$
and to display  the necessary commutation relations.  The energy-momentum
tensor of the non-interacting fields is defined as $T_0^{\mu\nu}$.  Use of
Eq.~(\ref{tmunu})  leads to the result
\begin{eqnarray}
T^{\mu\nu}_0 &=&
\partial^\mu \phi \partial^\nu \phi - \frac{g^{\mu\nu}}{2}
\left[\partial_\sigma \phi \partial^\sigma \phi-M^2\phi^2\right] 
+ \partial^\mu \chi \partial^\nu \chi - \frac{g^{\mu\nu}}{2}
\left[\partial_\sigma \chi \partial^\sigma \chi-\mu^2\chi^2\right],
\end{eqnarray}
with
\begin{eqnarray}
T^{+-}_0 
= \bbox{\nabla}_\perp \phi \cdot \bbox{\nabla}_\perp\phi + M^2 \phi^2 +
\bbox{\nabla}_\perp \chi \cdot \bbox{\nabla}_\perp\chi + \mu^2 \chi^2.
\end{eqnarray}

The scalar nucleon fields can be expressed in terms of creation and
destruction operators:
\begin{eqnarray}
\phi_i(x)&=&
\int \frac{d^2 k_\perp dk^+ \, \theta(k^+)}{(2\pi)^{3/2}\sqrt{2k^+}} \left[
a_i(\bbox{k})e^{-ik\cdot x} +a_i^\dagger(\bbox{k})e^{ik\cdot x}\right],
\end{eqnarray}
where $i=1,2$ is a particle index,
$k\cdot x={1\over2}(k^-x^++k^+x^-)-\bbox{k}_\perp \cdot \bbox{x}_\perp$ with
$k^-= \frac{M^2 + \bbox{k}_\perp^2}{k^+}$,
and $\bbox{k}\equiv(k^+,\bbox{k}_\perp)$.
Note that $k^-$ is such that the particles are on the mass shell, which
is a consequence of using a Hamiltonian theory.
The $\theta$ function restricts $k^+$ to positive values.
Likewise, the scalar meson field is given by
\begin{eqnarray}
\chi(x)&=&
\int \frac{d^2 k_\perp dk^+ \, \theta(k^+)}{(2\pi)^{3/2}\sqrt{2k^+}} \left[
a_\chi(\bbox{k})e^{-ik\cdot x} +a_\chi^\dagger(\bbox{k})e^{ik\cdot x}\right],
\end{eqnarray}
where $k^-= \frac{\mu^2 + \bbox{k}_\perp^2}{k^+}$,
so that the mesons are also on the
mass shell.  The non-vanishing commutation relations  are
\begin{eqnarray}
\left[a_\alpha(\bbox{k}),a_\alpha^\dagger(\bbox{k}')\right] &=&
\delta(\bbox{k}_\perp-\bbox{k}'_\perp)
\delta(k^+-k'^+), \label{comm}
\end{eqnarray}
where $\alpha = 1,2,\chi$ is a particle index.
The commutation relations are defined at equal light-front
time, $x^+=0$.  It is useful to define
\begin{eqnarray}
\delta^{(2,+)}(\bbox{k}-\bbox{k}') &\equiv&
\delta(\bbox{k}_\perp-\bbox{k}'_\perp)
\delta(k^+-k'^+),
\end{eqnarray}
which will be used throughout this paper.

We write a ket in the two-distinguishable-particle
sector of the Fock space as 
\begin{eqnarray}
|k_1,k_2 \rangle &=& a_1^\dagger(k_1) a_2^\dagger(k_2) |0\rangle.
\end{eqnarray}
This implies that the identity operator in this Fock space sector can be
written as
\begin{eqnarray}
I_{2} &=& \int d^2k_{1,\perp}dk_1^+ \int d^2k_{2,\perp}dk_2^+
|k_1,k_2 \rangle \langle k_1,k_2 | \label{twopartident}.
\end{eqnarray}

The derivatives appearing in the quantity $T^{+-}_0$ are evaluated and then one
sets $x^+$ to 0 to obtain the result
\begin{eqnarray}
P^-_0 &=& 
\int_k \, \left[
\frac{M^2 + \bbox{k}_\perp^2}{k^+}
\left( a_1^\dagger(k) a_1(k) + a_2^\dagger(k) a_2(k) \right) 
+ \frac{\mu^2 + \bbox{k}_\perp^2}{k^+}
a_\chi^\dagger(k) a_\chi(k)
\right], \label{p0minusop}
\end{eqnarray}
with $\int_k = \int d^2 k_\perp dk^+ \, \theta(k^+)$.  Eq.~(\ref{p0minusop})
has the interpretation of an operator that counts the light-front
energy $k^-$ (which is $\frac{M^2 + \bbox{k}_\perp^2}{k^+}$ for the 
nucleons and $\frac{\mu^2 + \bbox{k}_\perp^2}{k^+}$ for the mesons)
of all of the particles.  

We now consider the interacting part of the Lagrangian, ${\mathcal L}_I$.
An analysis similar to that for the non-interacting parts
yields the interacting part of the light-front
Hamiltonian $P_I^-$;
\begin{eqnarray}
P_I^- 
&=& 
\sum_{i=1,2} \frac{M}{2}
\int_k
\int_{k'}
\frac{1}{(2\pi)^{3/2} \sqrt{2 k^+ k'^+(k^++k'^+)} }
\nonumber\\ 
&& \phantom{\sum_{i=1,2} \frac{gM}{2} \int_k \int_{k'}}
\times \left\{ \left[
2 a_i^\dagger(k+k') a_\chi(k') a_i(k   )
+ a_\chi^\dagger(k+k') a_i(k') a_i(k   )
\right] \right. \nonumber\\
&& \phantom{\sum_{i=1,2} \frac{gM}{2} \int_k \int_{k'}
\times \left\{ \right. }
+ \mbox{Hermitian conjugate} \Big\}. 
\phantom{ \left.\right\} } \label{intham}
\end{eqnarray}
The interaction Hamiltonian is self-adjoint since the Hilbert space is
the Fock space.
The total light-front Hamiltonian is given by $P^- = P^-_0 + g P^-_I$.

\subsection{Hamiltonian bound-state equations}

We will be studying the bound states of two distinguishable nucleons.
The technology of time-ordered (old-fashioned) perturbation theory is
used to construct the light-front time-ordered perturbation theory (LFTOPT)
for our Hamiltonian.  We start with the light-front Schr\"odinger
equation in the full Fock space, 
\begin{eqnarray}
\left( P_0^- + g P_I^- \right)
| \psi^{\text{GS}}_F \rangle &=&
| \psi^{\text{GS}}_F \rangle
P^-_{\text{GS}}, \label{fullfock}
\end{eqnarray}
where $P_0^- + g P_I^-$ is the Hamiltonian in the full Fock-space basis,
$|\psi^{\text{GS}}_F\rangle$ is the ground-state wavefunction in the
full Fock space, and $P^-_{\text{GS}}$ is the light-front energy of that
state.  Recall that $P^-_0$, the non-interacting part of the
Hamiltonian, is diagonal in the momentum basis, while $P^-_I$, which
contains the interaction, has only off-diagonal elements.

A serious drawback of this equation is that the wavefunction
$\psi^{\text{GS}}_F \rangle$ has support from infinitely many sectors of
the Fock space, since $P^-_I$ changes the total number of particles.
However, the components of the wavefunction with many particles will be
small compared to the two-particle component if the coupling constant is
not too large.  We will construct the two-particle light-front
Schr\"odinger equation which the two-particle component of the
wavefunction satisfies.  From this construction, we will obtain the
rules for the LFTOPT.

We start by introducing the projection operators ${\mathcal P}$ and
${\mathcal Q}$.  The operator ${\mathcal P}$ projects out the sector of
Fock space with two distinguishable nucleons and no mesons, while
${\mathcal Q}=I-{\mathcal P}$ projects out all the other sectors.  We
define
\begin{eqnarray}
{\mathcal P} | \psi^{\text{GS}}_F \rangle & \equiv & | \psi^{\text{GS}}
\rangle \\
{\mathcal Q} | \psi^{\text{GS}}_F \rangle & \equiv & |
\psi^{\text{GS}}_Q \rangle, 
\end{eqnarray}
so that
$|\psi^{\text{GS}}_F\rangle=|\psi^{\text{GS}}\rangle+|\psi^{\text{GS}}_Q\rangle$.
Since the free Hamiltonian does not change the number of particles, 
$[{\mathcal P},P_0^-]=[{\mathcal Q},P_0^-]=0$.  The interaction
Hamiltonian changes the particle number, so it cannot
connect the two-particle sector to itself, thus
${\mathcal P} P_I^- {\mathcal P} = 0$. 

Using these projection operators, Eq.~(\ref{fullfock}) can be broken up
into two parts,
\begin{eqnarray}
P_0^- | \psi^{\text{GS}} \rangle 
+ g {\mathcal P} P_I^- {\mathcal Q} | \psi^{\text{GS}}_Q \rangle &=& |
\psi^{\text{GS}} \rangle P^-_{\text{GS}} \\ 
\left( P_0^- + g {\mathcal Q} P_I^- {\mathcal Q} \right) |
\psi^{\text{GS}}_Q \rangle 
+ g {\mathcal Q} P_I^- {\mathcal P} | \psi^{\text{GS}} \rangle &=& |
\psi^{\text{GS}}_Q \rangle P^-_{\text{GS}}. 
\end{eqnarray}
Eliminating the $|\psi^{\text{GS}}_Q \rangle$ and using the expression
of the identity given in Eq.~(\ref{twopartident}) we obtain the
two-particle effective light-front Schr\"odinger equation
\begin{eqnarray}
& &\int d^2p_{1,\perp} dp_1^+ \int d^2p_{2,\perp} dp_2^+
\langle \bbox{k}_1, \bbox{k}_2 |
\left[ P_0^- + V(g,P^-_{\text{GS}}) \right] 
| \bbox{p}_1, \bbox{p}_2 \rangle
\langle \bbox{p}_1, \bbox{p}_2 | \psi^{\text{GS}} \rangle \nonumber \\
& & \quad =
\langle \bbox{k}_1, \bbox{k}_2 | \psi^{\text{GS}} \rangle P^-_{\text{GS}},
\label{projfock}
\end{eqnarray}
where $P_0^-$ and the potential $V$ act in the two-nucleon basis.  The
two-particle potential is given by
\begin{eqnarray}
V(g,P^-) &=&  g^2 {\mathcal P} P_I^- \frac{{\mathcal Q}}{P^- - P_0^-
- g {\mathcal Q} P_I^- {\mathcal Q}} P_I^- {\mathcal P}.
\end{eqnarray}
Note that Eq.~(\ref{projfock}) is similar to Eq.~(\ref{fullfock}),
except for two main differences.  Here we have a two-nucleon
wavefunction, which makes it simpler.  However, the potential is
light-front energy dependent, which makes it more complicated. 

The denominator in the definition of the potential is non-diagonal
in the full Fock space, so the matrix inversion that it represents is
highly non-trivial.  This problem is avoided by expanding the inversion
in powers of the coupling constant $g$ to get
\begin{eqnarray}
V(g,P^-) &=& {\mathcal P} P_I^- \left[
\frac{g^2 {\mathcal Q}}{P^- - P_0^-} \sum_{n=0}^\infty 
\left( P_I^- \frac{g {\mathcal Q}}{P^- - P_0^-} \right)^n \right]
P_I^- {\mathcal P}.
\end{eqnarray}
This can be simplified further by noting that in the two-nucleon sector
of our theory, every meson
emitted must be absorbed, so there must be an even number of
interactions.  Thus, the full potential can be written as the sum of $n$
meson exchange potentials,
\begin{eqnarray}
V(P^-,g) &=& \sum_{n=1}^\infty g^{2n} V_{(2n)}(P^-), \label{fullpot}
\end{eqnarray}
where $V_{(2n)}$ is the potential due to the exchange of $n$ mesons,
given by
\begin{eqnarray}
V_{(2n)}(P^-) &=& {\mathcal P}
\left( P_I^- \frac{{\mathcal Q}}{P^- - P_0^-} \right)^{2n-1} P_I^-
{\mathcal P}.  \label{LFTOPpotdet}
\end{eqnarray}

It is easy to see how to write a sum of diagrams for the potential when
one says what Eq.~(\ref{LFTOPpotdet}) represents in words.
We start off with two
particles, then the interaction occurs.  There are two possibilities of
what can happen; nucleon 1 or 2 can emit a meson.  Each possibility has
a separate diagram.  After the interaction, there is propagation with
the light-front Green's function,
\begin{eqnarray}
G_{\text{LF}}(P^-) &=& \frac{1}{P^--P_0^-},
\end{eqnarray}
until another interaction occurs, and so on.  We simply sum up all of
the possible orderings of the interaction to get the full potential.
The $n^{\text{th}}$ order potential is simply the sum of all
possible diagrams with $n$-meson exchanges.

Each intermediate state in Eq.~(\ref{LFTOPpotdet}) has more than two
particles, so the diagrams are two-particle irreducible with respect to
the two-particle Green's function $G_{2\text{LF}}={\mathcal
P}G_{\text{LF}}{\mathcal P}$.  We can represent $G_{2\text{LF}}$ by its
diagonal matrix elements,
\begin{eqnarray}
G_{2\text{LF}}(\bbox{k}_1,\bbox{k}_2;P^-) = \frac{1}{P^- - k_1^- - k_2^-}.
\end{eqnarray}

In the diagrams we draw, the nucleons will be represented by solid lines
and the mesons by the dashed lines.  Although the states we will be
considering consist of two distinguishable nucleons, we will not
label the nucleon lines.  We will be using the quenched
approximation (so there are no nucleon loops) and neglect the mass and
vertex renormalization diagrams (so the physical masses and coupling
constant are used, and each meson emitted from one nucleon must
be absorbed by the other nucleon).  
It is not expected that these restrictions will lead to qualitatively
different results than the true full solution when the states are not
too deeply bound.  The quenched approximation is reasonable when the
masses of the nucleon fields are large compared to the binding energy.
Use of the physical masses and coupling constant are reasonable as well
when the momenta are not too large.

We stress again that we will compare various truncations of the
light-front Hamiltonian to other calculations which do not include
renormalization diagrams.  This is because we want to determine the
effect of truncation on the light-front Hamiltonian.  The differences
between our calculation and those which include the self-energy graphs
\cite{Wivoda:1993qr}, which may be large for deeply-bound states, are
not considered here.

The rules for drawing the $n$-meson-exchange graphs that
correspond to this approximation are:
\begin{enumerate}
\item{} Draw all topologically distinct time-ordered diagrams with $n$
mesons.  Use solid lines for the nucleons and dashed lines for the
mesons.
\item{} Delete all graphs which couple particles to the vacuum.  In the
massive theory we consider here, these diagrams always vanish since the
vacuum has zero plus momentum, and massive particles always have
positive plus momentum.
\item{} Our quenched approximation and use of the physical masses and
coupling constant requires us to delete all graphs that have nucleon
loops or have mesons that are emitted and absorbed from the same
nucleon.
\item{} Delete all other graphs which are not allowed in the particular
approximation that is being considered.  For example, consider the
potential from the the Hamiltonian theory that can be obtained from the
ladder Bethe-Salpeter equation.  That potential will not have any graphs
where the meson lines cross.
\end{enumerate}

Once the diagrams are drawn, we use the following rules to convert the
sum of diagrams into the potential
$\langle\bbox{k}_1,\bbox{k}_2|V_{(2n)}(P^-)|\bbox{p}_1,\bbox{p}_2\rangle$:
\begin{enumerate}
\item{} Overall factor of
$\frac{\delta^{(2,+)}(\bbox{k}_1+\bbox{k}_2-\bbox{p}_1-\bbox{p}_2)}{
2(2\pi)^3 \sqrt{k_1^+k_2^+p_1^+p_2^+}}$.  This delta function says that
the total light-front three-momentum is conserved.  We define the
light-front three-momentum $\bbox{P}\equiv\bbox{k}_1+\bbox{k}_2$. 
\item{} To each internal line, assign a light-front three-momentum
$\bbox{q}_i$ where $i=1,2,\ldots,N$ and $N$ is the number of internal lines.
The light-front energy for particle $i$ with mass $m_i$ is 
$q_i=\frac{m_1^2+\bbox{q}^2_{i,\perp}}{q_i^+}$.  It is useful to define
$z_i=q_i^+/P^+$.
\item{} A factor of $\frac{\theta(z_i^+)}{z_i^+}$ for each internal
line.
\item{} An extra factor of $\frac{M^2}{P^+P^-}$ for each internal meson
line.
\item{} A factor of $\frac{P^-}{\left( P^- - \sum_i q_i^-\right)}$
between consecutive vertices, where the sum is over only the particles
that exist in the intermediate time between those vertices.
\item{} Use light-front three-momentum conservation to eliminate all the
independent momenta.
\item{} Integrate with 
$\int \frac{d^2q_{i,\perp} dz_i}{2 P^+P^- (2\pi)^3}$ over all remaining
free internal momenta.
\item{} Symmetry factor of $\frac{1}{2}$ when two nucleons are created
or destroyed at the same time.
\end{enumerate} \label{rules}
With these rules, one can calculate the effective potential for any
order.

\subsection{Further development of the light-front Schr\"odinger equation}

Once the potential is calculated, we can plug it into
Eq.~(\ref{projfock}), which we as
\begin{eqnarray}
& &\int d^2p_{1,\perp} dp_1^+ \int d^2p_{2,\perp} dp_2^+
\langle \bbox{k}_1, \bbox{k}_2 |
\left[ P_0^- + V(g(P^-),P^-) \right] 
| \bbox{p}_1, \bbox{p}_2 \rangle
\langle \bbox{p}_1, \bbox{p}_2 | \psi^{\text{GS}} \rangle \nonumber \\
& & \quad =
\langle \bbox{k}_1, \bbox{k}_2 | \psi^{\text{GS}} \rangle P^-,
\label{fullse}
\end{eqnarray}
where $P^-$ is an arbitrary light-front energy and $g(P^-)$ is the
coupling constant which yields the bound-state wavefunction with $P^-$
as the bound-state energy.  We call this $g(P^-)$ the spectrum of the
light-front Schr\"odinger equation for the corresponding wavefunction.

The total momentum $\bbox{P}=\bbox{k}_1+\bbox{k}_2$ is conserved by the
potential given in Eq.~(\ref{fullpot}), so the wavefunction in
Eq.~(\ref{fullse}) can be parameterized by the total momentum.  To make
the calculations easier later, we choose to be in the center-of-momentum
frame, where the components of the total momentum can be written as
$\bbox{P}_\perp=0$ and $P^+=P^-=E$.  The ground-state energy, $E$, is
the same as the mass of the bound state.   In terms of the binding
energy $B$, $E=2M-B$.  In the center-of-momentum frame, the ground-state
wavefunction is parameterized by $E$, so we can define
\begin{eqnarray}
\langle \bbox{k}_1,\bbox{k}_2 |\psi^{\text{GS}}_M\rangle
&=&
\delta^{(2,+)}(\bbox{k}_1+\bbox{k}_2-\bbox{P})
\psi^{\text{GS}}(\bbox{k}_1) \label{twopartredpsi} \\
\langle \bbox{k}_1,\bbox{k}_2| V_{(2n)}(P^-) | \bbox{p}_1,\bbox{p}_2 \rangle
&=& \delta^{(2,+)}(\bbox{k}_1+\bbox{k}_2-\bbox{p}_1-\bbox{p}_2)
V_{(2n)}(E;\bbox{k}_1;\bbox{p}_1) \label{twopartredpot}.
\end{eqnarray}
With these, Eq.~(\ref{fullse}) effectively becomes a one-particle
equation, where particle 2's momentum is determined by
$\bbox{k}_2=\bbox{P}-\bbox{k}_1$.  The minus component (the light-front
energy) of particle 2 is defined by the requirement that the particle 2
is on mass shell, so $k^-_2=(M^2+\bbox{k}_{2,\perp}^2)/k_2^+$. We also
define $x \equiv k_1^+/P^+ = x_{Bj}$, where $x_{Bj}$ is the Bjorken $x$
variable, so that $k_2^+/P^+=1-x$.  Likewise, we write the Bjorken
variables that correspond to the momenta $\bbox{p}_1$ and $\bbox{q}_1$
as $y\equiv p_1^+/P^+$ and $z\equiv q_1^+/P^+$.

Using Eqs.~(\ref{twopartredpsi}), (\ref{twopartredpot}), and
the fact that the plus momentum of both nucleons is positive, we can
write the light-front Schr\"odinger equation Eq.~(\ref{fullse}) as
\begin{eqnarray}
\int d^2p_{1,\perp} \int_0^E dp^+_1 V(g(E),E;\bbox{k}_1;\bbox{p}_1)
\psi^{\text{GS}}(\bbox{p}_1)
&=& 
\psi^{\text{GS}}(\bbox{k}_1) (E-k_1^--k_2^-). \label{fullse3}
\end{eqnarray}

It useful to convert from light-front coordinates
$\bbox{k}_1=(k_1^+,\bbox{k}_\perp)$ to equal-time coordinates
$\bbox{k}_{\text{ET}}=( \bbox{k}_\perp,k^3)$, using an implicit
definition of $k^3$ 
\cite{Terentev:1976jk}
\begin{eqnarray}
k_1^+ &=& \frac{E}{2k^0(\bbox{k}_{\text{ET}})}
\left[ k^0(\bbox{k}_{\text{ET}}) + k^3 \right] \label{eteq} \\
k^0(\bbox{k}_{\text{ET}}) &=& \sqrt{ M^2 + \bbox{k}_{\text{ET}}^2 }.
\end{eqnarray}
Often the explicit dependence of $k^0$ on $\bbox{k}_{\text{ET}}$ will
not be shown.  It is worth emphasizing that this is just a convenient
change of variables; $\psi^{\text{GS}}(\bbox{k}_{\text{ET}})$ is not the
usual equal-time ground-state wavefunction.  With this transformation,
we can express $k_1^\pm$ and $k_2^\pm$ as 
\begin{mathletters}
\label{pminet}
\begin{equation}
k_1^+ = k_2^- \left( \frac{E}{2k^0} \right)^2 =
\left(1+\frac{k^3}{k^0}\right)\frac{E}{2},
\end{equation}
\begin{equation}
k_2^+ = k_1^- \left( \frac{E}{2k^0} \right)^2 =
\left(1-\frac{k^3}{k^0}\right)\frac{E}{2}.
\end{equation}
\end{mathletters}
Using these, Eq.~(\ref{fullse3}) becomes
\begin{eqnarray}
\int d^3p_{\text{ET}}
\frac{2 p_1^+ p_2^+}{p^0}
V(g(E),E,\bbox{k}_{\text{ET}};\bbox{p}_{\text{ET}})
\psi^{\text{GS}}(\bbox{p}_{\text{ET}})
&=& 
\psi^{\text{GS}}(\bbox{k}_{\text{ET}})
\left[E^2 - (2k^0)^2 \right].
\label{fullse4}
\end{eqnarray}

Now consider the exchange of the particle labels 1 and 2.  This causes
\begin{eqnarray}
\bbox{k}_{1,\perp} &\rightarrow& \bbox{k}_{2,\perp} = -\bbox{k}_{1,\perp} \\
k_1^+ &\rightarrow& k_2^+ = E-k_1^+,
\end{eqnarray}
which means that $k^3$ as defined in Eq.~(\ref{eteq}) transforms as
$k^3 \rightarrow -k^3$, so $\bbox{k}_{\text{ET}} \rightarrow
-\bbox{k}_{\text{ET}}$.  Consequently, exchange of particle labels 1 and
2 is the same as parity for $\bbox{k}_{\text{ET}}$.

Since the two nucleons are identical except for the particle label, the
effective potential commutes with parity to all orders in $g^2$. Furthermore,
the light-front Hamiltonian is explicitly invariant under rotations about the
three-axis. These considerations allow us to classify the wavefunctions
as eigenfunctions of parity and the three-component of the angular
momentum operator.  The ground state will have even parity and be
invariant under rotation about the three-axis, so we can write
\begin{eqnarray}
\psi^{\text{GS}}(\bbox{k}_{\text{ET}}) =
\psi^{\text{GS}}(k_{\text{ET}},\theta_k,\phi_k) =
\psi^{\text{GS}}(k_{\text{ET}},    \theta_k) =
\psi^{\text{GS}}(k_{\text{ET}},\pi-\theta_k),
\end{eqnarray}
where $\theta_k$ and $\phi_k$ are the polar and azimuthal angles,
respectively, for the vector $\bbox{k}_{\text{ET}}$.

The cylindrical symmetry and parity of the wavefunction can be used to
rewrite Eq.~(\ref{fullse4}) as
\begin{eqnarray}
& & \int_0^\infty dp_{\text{ET}} \int_0^{\pi/2} d\theta_p
\frac{2 p_1^+ p_2^+  p^2_{\text{ET}} \sin\theta_p}{p^0}
V^+(k_{\text{ET}},\theta_k;p_{\text{ET}},\theta_p)
\psi^{\text{GS}}(p_{\text{ET}},\theta_p)
\nonumber \\ & & \qquad = 
\psi^{\text{GS}}(k_{\text{ET}},\theta_k)
\left[E^2 - 4 (k^0)^2 \right],
\label{fullse6}
\end{eqnarray}
where
\begin{eqnarray}
V^+(k_{\text{ET}},\theta_k;p_{\text{ET}},\theta_p)
&=& \frac{1}{2} \Big[
V(k_{\text{ET}},\theta_k;p_{\text{ET}},\theta_p) +
V(k_{\text{ET}},\theta_k;p_{\text{ET}},\pi-\theta_p) \Big]\\
V(k_{\text{ET}},\theta_k;p_{\text{ET}},\theta_p)
&=& \int \! \! \!\int_0^{2\pi} \frac{d\phi_k\,d\phi_p}{2\pi}
V(\bbox{k}_{\text{ET}};\bbox{p}_{\text{ET}}). \label{angintet}
\end{eqnarray}
We call $V(k_{\text{ET}},\theta_k;p_{\text{ET}},\theta_p)$ the
azimuthal-angle-averaged potential.
The light-front vectors can also be expressed in terms of the azimuthal 
angle $\phi$, where $\bbox{k}=(k^+,k_\perp,\phi_k)$, which allows the
azimuthal-angle-averaged potential to be written in light-front coordinates
\begin{eqnarray}
V(k^+_1,k_{1,\perp};p_1^+,p_{1,\perp}) &=& \int\!\!\!\int_0^{2\pi}
\frac{d\phi_k\,d\phi_p}{2\pi} V(\bbox{k}_1;\bbox{p}_1). \label{angintlf}
\end{eqnarray}

All of the simplifications of Eq.~(\ref{fullse6}) based on physical
considerations have been addressed.  However, further rearrangements need
to be done before Eq.~(\ref{fullse6}) is fit to be solved on the
computer.  Since these involve only numerical techniques, they are
relegated to Appendix \ref{numericaleq}.

\section{Perturbative Potentials}
\label{apptrunc}

A truncation must be made of the expansion of the potential given in
Eq.~(\ref{fullpot}), since it is not feasible to calculate the infinite
sum of graphs for the potential in this Hamiltonian theory.
In this paper we consider three truncations of the
potential derived from the field theory.  First the OBE potential and the
TBE potentials are calculated.  Then, we note that a subset of the TBE
diagrams, the stretched-box diagrams, correspond to the truncated
potential derived from the ladder Bethe-Salpeter equation.  Thus,
three truncated potentials are obtained that have a physical
interpretation.

The matrix elements of these potentials are written in the two-particle
momentum basis, denoting the momentum of the incoming particles by
$\bbox{p}_1$ and $\bbox{p}_2$, and the outgoing particles $\bbox{k}_1$
and $\bbox{k}_2$.  For simplicity, we
choose to work in the center-of-momentum frame.  By inspecting the rules for
converting a light-front time-ordered diagram into a potential given in
section \ref{rules}, and looking at Eq.~(\ref{twopartredpot}), we find
that each piece of the effective-one-particle potential
$V(E;\bbox{k}_1,\bbox{p}_1)$ is proportional to
\begin{eqnarray}
\frac{E}{2(2\pi)^3 \sqrt{k_1^+k_2^+p_1^+p_2^+}}. \label{supressfactor}
\end{eqnarray}
(An extra factor of $E$ is included to simplify later equations.)
This term will by suppressed in all of the potentials written in this
paper.

\subsection{OBE Potential}

\begin{figure}[!ht]
\begin{center}
\epsfig{angle=0,width=2.732in,height=1.173in,file=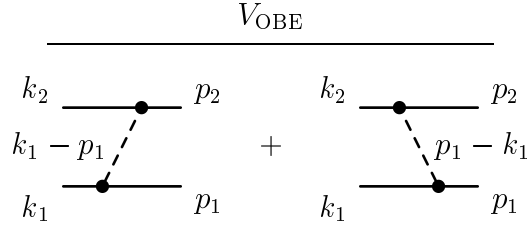}
\caption{The diagrams for the OBE potential.
\label{obediagrams}}
\end{center}
\end{figure}

We start by drawing all the allowed and non-vanishing time-ordered
diagrams with one meson exchange.  These diagrams are shown in 
Fig.~\ref{obediagrams}.
The light-front time-ordered perturbation theory rules given
in section \ref{rules} are used to calculate the potential due to OBE
potential,
\begin{eqnarray}
V_{\text{OBE}}(E;\bbox{k}_1;\bbox{p}_1)
&=& \left(\frac{M}{E}\right)^2 \left[
\frac{ \theta(x-y)/|x-y|}{E - p^-_1 - k^-_2 - \omega^-(\bbox{k}_1-\bbox{p}_1)}
\right. \nonumber \\
&& 
\phantom{\left(\frac{M}{E}\right)^2 \left[ \right.} \left. 
+
\frac{ \theta(y-x)/|y-x|}{E - k^-_1 - p^-_2 - \omega^-(\bbox{p}_1-\bbox{k}_1)}
\right] \label{OBEpot}.
\end{eqnarray}
We have introduced the notation that meson with light-front
three-momentum $\bbox{q}$ has a light-front energy given by
\begin{eqnarray}
\omega^-(\bbox{q}) &=& \frac{\mu^2 + \bbox{q}_\perp^2}{q^+}.
\end{eqnarray}
The azimuthal-angle average of $V_{\text{OBE}}$ is discussed in Appendix
\ref{angint}.

The potential given in Eq.~(\ref{OBEpot}) can also be used for
scattering states.  In that case, $E=k_1^-+k_2^-=p_1^-+p_2^-$, which 
allows the potential to be written as 
\begin{eqnarray}
V_{\text{OBE}}(E_{\text{scat}};\bbox{k}_1;\bbox{p}_1)
&=& \frac{M^2/E_{\text{scat}}}{(k_1-p_1)^2 - \mu^2 }.
\label{OBEpotscat}
\end{eqnarray}
The scattering potential is the same as the usual equal-time OBE
potential.  This must be the case, since the scattering potential is
also given by covariant Feynman diagrams, which have the same form
independent of the form of dynamics.

Returning to the bound-state regime, we note that the OBE potential can
easily be written in terms of the equal-time coordinates.  A
reorganization of Eq.~(\ref{OBEpot}) yields
\begin{eqnarray}
V_{\text{OBE}}(E;\bbox{k}_{\text{ET}};\bbox{p}_{\text{ET}})
&=& \left(\frac{M}{E}\right)^2 E \left[
\frac{\theta(x-y)}
{(k_1^+-p_1^+)(E - p^-_1 - k^-_2) - \mu^2 - (\bbox{p}_\perp-\bbox{k}_\perp)^2}
\right. \nonumber \\
&&
\phantom{\left(\frac{M}{E}\right)^2 E \left[ \right.} \left. 
+
\frac{\theta(y-x)}
{(p_1^+-k_1^+)(E - k^-_1 - p^-_2) - \mu^2 - (\bbox{p}_\perp-\bbox{k}_\perp)^2}
\right]. \label{OBEpot2}
\end{eqnarray}
Using the relations in Eq.~(\ref{pminet}), we find
\begin{eqnarray}
(k_1^+-p_1^+)(E - p^-_1 - k^-_2)
&=&
\left(\frac{k^3}{k^0}-\frac{p^3}{p^0}\right)
\left(\frac{E^2-4M^2}{2} -p_{\text{ET}}^2 -k_{\text{ET}}^2 \right)
\nonumber \\ & & 
+ \frac{k^3}{k^0} \frac{p^3}{p^0} \left(k^0-p^0\right)^2
- (k^3-p^3)^2. \label{etred}
\end{eqnarray}
Under the exchange $k\leftrightarrow p$, the only thing that changes in
Eq.~(\ref{etred}) is that the first term picks up a minus sign.
This observation allows Eq.~(\ref{OBEpot2}) to be rewritten as
\begin{eqnarray}
V_{\text{OBE}}(E;\bbox{k}_{\text{ET}};\bbox{p}_{\text{ET}})
&=& \left(\frac{M}{E}\right)^2
\frac{E}
{\left|\frac{k^3}{k^0}-\frac{p^3}{p^0}\right|
\Delta
+ \frac{k^3}{k^0} \frac{p^3}{p^0} (q^0_{\text{ET}})^2 
- \bbox{q}_{\text{ET}}^2
- \mu^2
},
\label{OBEpotet}
\end{eqnarray}
where
\begin{eqnarray}
\Delta &=& \frac{E^2-4M^2}{2} -p_{\text{ET}}^2 -k_{\text{ET}}^2 \\
q_{\text{ET}}^\mu &=& k^\mu_{\text{ET}} - p^\mu_{\text{ET}}.
\end{eqnarray}
Note that $q_{\text{ET}}^-$ is not the light-front energy of the meson,
since in a Hamiltonian theory only the light-front three-momenta are
conserved; the four-momenta is not conserved.  Equation (\ref{OBEpotet}) 
will be useful in the context of approximations based on the physical
arguments that we will discuss in section \ref{appphys}.

\subsection{TBE Potential}
\label{deftbepots}

\begin{figure}[!ht]
\begin{center}
\epsfig{angle=0,width=4.278in,height=6.100in,file=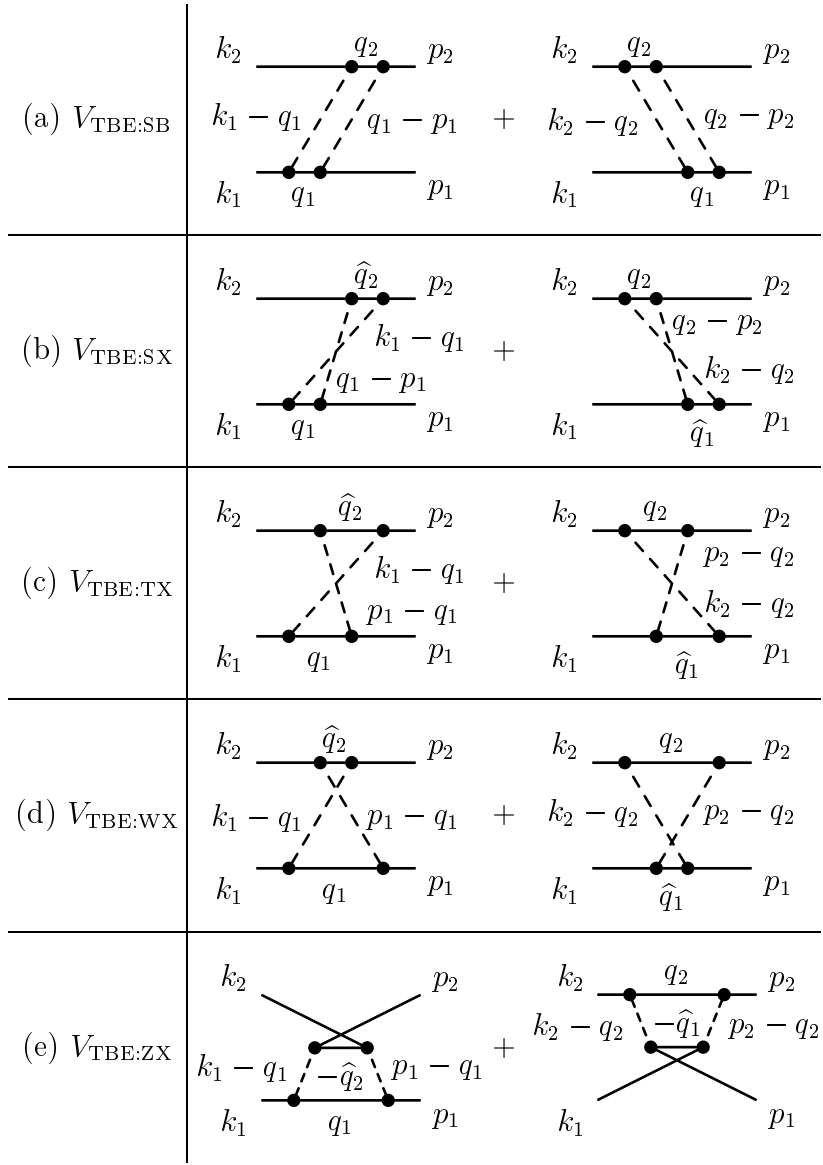}
\caption{The components of the TBE potential, (a) the stretched-box,
(b) stretched-crossed, (c) T-crossed, (d) wide-crossed, and (e)
Z-crossed diagrams.  Here, $\widehat{q}_1=k_1+p_1-q_1$
and $\widehat{q}_2=k_2+p_2-q_2$.
\label{tbediagrams}}
\end{center}
\end{figure}

As in the previous section, we start by drawing all the allowed,
non-vanishing time-ordered diagrams with two meson exchanges shown in
Fig.~\ref{tbediagrams}.  The diagrams are classified according to the
behavior of the intermediate particles.  The total TBE potential is
given by the sum of all the diagrams, so
\begin{eqnarray}
V_{\text{TBE}} &=& 
V_{\text{TBE:SB}} +
V_{\text{TBE:SX}} +
V_{\text{TBE:TX}} +
V_{\text{TBE:WX}} +
V_{\text{TBE:ZX}}.
\end{eqnarray}

In the diagrams for the TBE potential in Fig.~\ref{tbediagrams}, the
intermediate loop momenta can be parameterized by $\bbox{q}_1$ or
$\bbox{q}_2$.  The dependent variable is defined by the relation
$\bbox{P}=\bbox{q}_1+\bbox{q}_2$.  The Bjorken $x$ variable that
corresponds to $\bbox{q}_1$ ($\bbox{q}_2$) is labeled with $z$ ($1-z$).
We use the Feynman rules to calculate all of these potentials, starting
with
\begin{eqnarray}
V_{\text{TBE:SB}}(E;\bbox{k}_1;\bbox{p}_1) &=&
\left(\frac{M}{E}\right)^4
\int \frac{d^2 q_\perp}{2(2\pi)^3} \left[
\int_0^1 dz \frac{\theta(z-y)\theta(x-z)}{z(1-z)(z-y)(x-z)}
\right. \nonumber \\ && \phantom{\left(\frac{M}{E}\right)^4 \int
\frac{d^2 q_\perp}{2(2\pi)^3} \left[\right.} \left. \times
\frac{1}{E -q_1^- - k_2^- - \omega^-(\bbox{k}_1-\bbox{q}_1)}
\right. \nonumber \\ && \phantom{\left(\frac{M}{E}\right)^4 \int
\frac{d^2 q_\perp}{2(2\pi)^3} \left[\right.} \left. \times
\frac{1}{E -p_1^- - k_2^- - \omega^-(\bbox{k}_1-\bbox{q}_1) -
\omega^-(\bbox{q}_1-\bbox{p}_1)}
\right. \nonumber \\ && \phantom{\left(\frac{M}{E}\right)^4 \int
\frac{d^2 q_\perp}{2(2\pi)^3} \left[\right.} \left. \times
\frac{1}{E -p_1^- - q_2^- - \omega^-(\bbox{q}_1-\bbox{p}_1)} \right]
\nonumber \\ && \phantom{\left(\frac{M}{E}\right)^4 \int
\frac{d^2 q_\perp}{2(2\pi)^3}} 
+ \{ 1\leftrightarrow 2\} \label{TBESBpot}.
\end{eqnarray}
The symbol $\{ 1\leftrightarrow 2\}$ means that all labels 1 are replaced with
2 and vice versa, as well as replacing the Bjorken variables $x$, $y$,
and $z$ with $1-x$, $1-y$, and $1-z$.  This is a way of explicitly
stating the symmetry of the potential under exchange of particles 1 and
2.  A detailed discussion of the evaluation of the loop integral in
Eq.~(\ref{TBESBpot}) is given in Appendix \ref{loopints}. 

It is straightforward to calculate the other parts of the TBE potential,
\begin{eqnarray}
V_{\text{TBE:SX}}(E;\bbox{k}_1;\bbox{p}_1) &=&
\left(\frac{M}{E}\right)^4
\int \frac{d^2 q_\perp}{2(2\pi)^3} \left[
\int_0^1 dz \frac{\theta(x-z)\theta(z-y)}{z(x-z)(1+z-x-y)(z-y)}
\right. \nonumber \\ && \phantom{\left(\frac{M}{E}\right)^4 \int
\frac{d^2 q_\perp}{2(2\pi)^3} \left[\right.} \left. \times
\frac{1}{E -q_1^- - k_2^- - 
\omega^-(\bbox{k}_1-\bbox{q}_1)}
\right. \nonumber \\ && \phantom{\left(\frac{M}{E}\right)^4 \int
\frac{d^2 q_\perp}{2(2\pi)^3} \left[\right.} \left. \times
\frac{1}{E -p_1^- - k_2^- - 
\omega^-(\bbox{k}_1-\bbox{q}_1) -
\omega^-(\bbox{q}_1-\bbox{p}_1)}
\right. \nonumber \\ && \phantom{\left(\frac{M}{E}\right)^4 \int
\frac{d^2 q_\perp}{2(2\pi)^3} \left[\right.} \left. \times
\frac{1}{E -p_1^- - \widehat{q}_2^- -
\omega^-(\bbox{k}_1-\bbox{q}_1)}
\right]
\nonumber \\ && \phantom{\left(\frac{M}{E}\right)^4 \int
\frac{d^2 q_\perp}{2(2\pi)^3}} 
+ \{ 1\leftrightarrow 2\} \label{TBESXpot},
\end{eqnarray}
where we have denoted the light-front energy of particle 2 by
$\widehat{q}_2^-$, given by
\begin{eqnarray}
\widehat{q}_2^- &=&
\epsilon^-(\bbox{P}+\bbox{q}_1-\bbox{p}_1-\bbox{k}_1) \\
\epsilon^-(\bbox{q}) &=& \frac{M^2 + \bbox{q}_\perp^2}{q^+}.
\end{eqnarray}
The rest of the TBE potential is given by
\begin{eqnarray}
V_{\text{TBE:TX}}(E;\bbox{k}_1;\bbox{p}_1) &=&
\left(\frac{M}{E}\right)^4
\int \frac{d^2 q_\perp}{2(2\pi)^3} \left[
\int_0^1 dz \frac{
\theta(x-z)\theta(1+z-x-y)\theta(y-z)}{z(x-z)(1+z-x-y)(y-z)}
\right. \nonumber \\ && \phantom{\left(\frac{M}{E}\right)^4 \int
\frac{d^2 q_\perp}{2(2\pi)^3} \left[\right.} \left. \times
\frac{1}{E -q_1^- - k_2^- - 
\omega^-(\bbox{k}_1-\bbox{q}_1)}
\right. \nonumber \\ && \phantom{\left(\frac{M}{E}\right)^4 \int
\frac{d^2 q_\perp}{2(2\pi)^3} \left[\right.} \left. \times
\frac{1}{E -q_1^- - \widehat{q}_2^- -
\omega^-(\bbox{k}_1-\bbox{q}_1) -
\omega^-(\bbox{p}_1-\bbox{q}_1)}
\right. \nonumber \\ && \phantom{\left(\frac{M}{E}\right)^4 \int
\frac{d^2 q_\perp}{2(2\pi)^3} \left[\right.} \left. \times
\frac{1}{E -p_1^- - \widehat{q}_2^- -
\omega^-(\bbox{k}_1-\bbox{q}_1)}
\right]
\nonumber \\ && \phantom{\left(\frac{M}{E}\right)^4 \int
\frac{d^2 q_\perp}{2(2\pi)^3}} 
+ \{ 1\leftrightarrow 2\} \label{TBETXpot} \\
V_{\text{TBE:WX}}(E;\bbox{k}_1;\bbox{p}_1) &=&
\left(\frac{M}{E}\right)^4
\int \frac{d^2 q_\perp}{2(2\pi)^3} \left[
\int_0^1 dz \frac{\theta(x-z)\theta(1+z-x-y)\theta(y-z)}{z(x-z)(1+z-x-y)(y-z)}
\right. \nonumber \\ && \phantom{\left(\frac{M}{E}\right)^4 \int
\frac{d^2 q_\perp}{2(2\pi)^3} \left[\right.} \left. \times
\frac{1}{E -q_1^- - k_2^- -
\omega^-(\bbox{k}_1-\bbox{q}_1)}
\right. \nonumber \\ && \phantom{\left(\frac{M}{E}\right)^4 \int
\frac{d^2 q_\perp}{2(2\pi)^3} \left[\right.} \left. \times
\frac{1}{E -q_1^- - \widehat{q}_2^- -
\omega^-(\bbox{k}_1-\bbox{q}_1) -
\omega^-(\bbox{p}_1-\bbox{q}_1)}
\right. \nonumber \\ && \phantom{\left(\frac{M}{E}\right)^4 \int
\frac{d^2 q_\perp}{2(2\pi)^3} \left[\right.} \left. \times
\frac{1}{E - q_1^- - p_2^- -
\omega^-(\bbox{p}_1-\bbox{q}_1)}
\right]
\nonumber \\ && \phantom{\left(\frac{M}{E}\right)^4 \int
\frac{d^2 q_\perp}{2(2\pi)^3}} 
+ \{ 1\leftrightarrow 2\} \label{TBEWXpot} \\
V_{\text{TBE:ZX}}(E;\bbox{k}_1;\bbox{p}_1) &=&
\left(\frac{M}{E}\right)^4
\int \frac{d^2 q_\perp}{2(2\pi)^3} \left[
\int_0^1 dz \frac{\theta(x+y-1-z)}{z(x-z)(x+y-1-z)(y-z)}
\right. \nonumber \\ && \phantom{\left(\frac{M}{E}\right)^4 \int
\frac{d^2 q_\perp}{2(2\pi)^3} \left[\right.} \left. \times
\frac{1}{E -q_1^- - k_2^- -
\omega^-(\bbox{k}_1-\bbox{q}_1)}
\right. \nonumber \\ && \phantom{\left(\frac{M}{E}\right)^4 \int
\frac{d^2 q_\perp}{2(2\pi)^3} \left[\right.} \left. \times
\frac{1}{E - q_1^- - k_2^- - p_2^- -
\epsilon^-(\bbox{p}_1+\bbox{k}_1-\bbox{P}-\bbox{q}_1-)}
\right. \nonumber \\ && \phantom{\left(\frac{M}{E}\right)^4 \int
\frac{d^2 q_\perp}{2(2\pi)^3} \left[\right.} \left. \times
\frac{1}{E - q_1^- - p_2^- -
\omega^-(\bbox{p}_1-\bbox{q}_1)}
\right]
\nonumber \\ && \phantom{\left(\frac{M}{E}\right)^4 \int
\frac{d^2 q_\perp}{2(2\pi)^3}} 
+ \{ 1\leftrightarrow 2\} \label{TBEZXpot}.
\end{eqnarray}
The loop integrals in the expressions for the TBE potentials and the
azimuthal-angle averaging are discussed in Appendix \ref{loopints}.

\subsection{TBE:SB Potential: Connection to the ladder Bethe-Salpeter
equation}
\label{tbe:sbSec} 

It is well known that the full, untruncated Bethe-Salpeter equation can
be reduced to the full, untruncated Hamiltonian (Schr\"odinger-type)
equation by integration over the energy or light-front energy
variables.  If a truncated kernel is used for the Bethe-Salpeter
equation, then the physically equivalent Hamiltonian equation will not
include all the graphs that the full theory allows.  By physically
equivalent, we mean that the spectra of the potential $V$ should
reproduce the spectrum for the states of the Bethe-Salpeter equation,
excluding the so-called ``abnormal'' states \cite{WickCutkosky}.
For an extensive discussion of
this in the equal-time case see, for instance, Klein \cite{klein},
Phillips and Wallace \cite{Phillips:1996eb}, Lahiff and Afnan
\cite{Lahiff:1997bj}, and for examples on the light front, Chang and Ma
\cite{Chang:1969bh} and Ligterink and Bakker \cite{Ligterink:1995tm}. 

In particular, consider the Bethe-Salpeter equation when the
ladder kernel is used.  The physically equivalent light-front potential
will not 
include any graphs where the meson lines cross, so to order $g^4$, the
potential is given by $g^2 V_{\text{OBE}}+ g^4 V_{\text{TBE:SB}}$.
Therefore, by 
considering the TBE:SB truncation, we can test how well the light-front
Hamiltonian approach approximates the full ladder Bethe-Salpeter
equation.  This is idea discussed more throughly in
\cite{Cooke:1999yi,Sales:1999ec,Ligterink:1995tm}.

\section{Non-perturbative Potentials}
\label{appphys}

The potentials discussed in this section are derived from the OBE field
theory potential, but additional approximations are made to simplify
the expressions.

\subsection{Symmetrized-mass approximation}

Krautg\"artner, Pauli and W\"olz \cite{Krautgartner:1992xz} and
Trittmann and Pauli \cite{Trittmann:1997ga} studied positronium with a
large coupling constant in light-front dynamics.  The
one-photon-exchange potential they obtain has a colinear singularity due
to the sum of the instantaneous photon exchange graph and a
gauge-dependent factor from the spin sum.  They argue that the
singularity is not physical, and therefore must be canceled by
higher-order terms in the potential.  The effect of those terms can be
simulated by choosing the bound-state energy so that the coefficient of
the singular term vanishes.  They find that this condition is met when
the light-front energy $P^-$ in the one-photon-exchange potential is
replaced with the operator $\omega$, expressed here in the two-particle
basis,
\begin{eqnarray}
P^- \Rightarrow
\omega\left(\bbox{k}_1,\bbox{k}_2;\bbox{p}_1,\bbox{p}_2\right)
&\equiv&
\frac{1}{2} \left(p_1^- + p_2^- + k_1^- +k_2^- \right).
\end{eqnarray}
This is called the {\it symmetrized mass} \cite{Krautgartner:1992xz},
the average of the total $P^-$ in the initial and final states.  It is
important to note that this approximation affects not only the singular
term, but also the energy denominator in the rest of the OBE potential.
The modified denominators simulate the effects of the non-perturbative
higher-order terms that are not included explicitly in the OBE
potential.  Potentials 
obtained with this approximation are similar to those given by the
unitary transformation method \cite{Eden:1996ey}, where the potentials
depend explicitly on the initial- and final-state energies.

In our model, there are no singularities associated with the OBE graphs
because we deal only with scalar fields.  However, we may use their
approximation to obtain a new light-front OBE potential that should
incorporate some non-perturbative effects.  Recalling that the only
place where $P^-$ occurred in Eq.~(\ref{OBEpot}) was in the denominator,
the $E$ in the denominator of the OBE potential is replaced with
$\omega$ to get
\begin{eqnarray}
V_\omega(\bbox{k}_1;\bbox{p}_1)
&=& \left(\frac{M}{E}\right)^2 \left[
\frac{ \theta(x-y)/|x-y|}{
\frac{1}{2}\left(- p_1^- + p_2^- + k_1^- - k_2^- \right)
- \omega^-(\bbox{k}_1-\bbox{p}_1)}
\right. \nonumber \\
&& 
\phantom{\left(\frac{M}{E}\right)^2 \left[ \right.} \left. 
+
\frac{ \theta(y-x)/|y-x|}{
\frac{1}{2}\left(+ p_1^- - p_2^- - k_1^- + k_2^- \right)
- \omega^-(\bbox{p}_1-\bbox{k}_1)}
\right] \\
&=& \left(\frac{M}{E}\right)^2
\frac{E}{\frac{1}{2} (k_1^+-p_1^+)\left(- p_1^- + p_2^- + k_1^- - k_2^- \right)
- \mu^2 - (\bbox{k}_\perp-\bbox{p}_\perp)^2}.
\label{presymmmass}
\end{eqnarray}
Writing the light-front variables in the denominator in terms of the
equal-time variables, as prescribed in Eq.~(\ref{pminet}), we find
\begin{eqnarray}
\frac{1}{2}(k_1^+-p_1^+)\left(- p_1^- + p_2^- + k_1^- - k_2^- \right)
&=& -(k^3 - p^3)^2 + \frac{k^3}{k^0} \frac{p^3}{p^0} (k^0-p^0)^2.
\end{eqnarray}
Thus, Eq.~(\ref{presymmmass}) can be rewritten as
\begin{eqnarray}
V_\omega(\bbox{k}_{\text{ET}};\bbox{p}_{\text{ET}})
&=& \left(\frac{M}{E}\right)^2
\frac{E}{
\frac{k^3}{k^0} \frac{p^3}{p^0} (k^0-p^0)^2
 - (\bbox{k}_{\text{ET}}-\bbox{p}_{\text{ET}})^2 - \mu^2}
\label{symmmass}.
\end{eqnarray}

This result can also by obtained more directly by considering the first
term in Eq.~(\ref{etred}).  Recall that the $E^2$ that appears in the
denominator is written as  $P^+P^-$ in an arbitrary frame, so in the
symmetrized-mass approximation, the $E^2$ term is replaced with
$E\omega$.  This causes $\Delta$ term in the denominator of 
Eq.~(\ref{OBEpotet}) to vanish, so the equation reduces to
Eq.~(\ref{symmmass}).  Also, note that by writing this new potential,
we attempt to incorporate physics from higher-order graphs than just the
OBE graphs.

The singularity structure of the symmetrized-mass potential is easily
analyzed. When 
scattering states are used, in the center-of-momentum frame the total
energy of the state is $E=2k^0=2p^0$, so the relations in
Eq.~(\ref{pminet}) become
\begin{mathletters}
\begin{equation}
k_1^\pm = k^0 \pm k^3,
\end{equation}
\begin{equation}
k_2^\pm = k^0 \mp k^3.
\end{equation}
\end{mathletters}
Using these relations, the symmetrized mass is $\omega = E$.  Thus, for
scattering states, this potential is same as the OBE scattering
potential and the singularity structure is the same.

\subsection{Instantaneous and Retarded approximations}

For our bound states, $k_{\text{ET}}^2 \ll M^2$, so that the
$\frac{k^3 p^3}{k^0 p^0}(k^0-p^0)^2$ will be much smaller than the other
terms in the denominator of  Eq.~(\ref{symmmass}).  Therefore, we may
approximate the symmetrized-mass potential $V_\omega$ by the
instantaneous potential,
\begin{eqnarray}
V_{\text{Inst}}(\bbox{k}_{\text{ET}};\bbox{p}_{\text{ET}})
&=& \left(\frac{M}{E}\right)^2
\frac{-E}{(\bbox{k}_{\text{ET}}-\bbox{p}_{\text{ET}})^2 + \mu^2}
\label{instpot}.
\end{eqnarray}
Alternatively, we may also argue that since the energy difference term
is small, we can also approximate $V_\omega$ by the retarded potential,
\begin{eqnarray}
V_{\text{Ret}}(\bbox{k}_{\text{ET}};\bbox{p}_{\text{ET}})
&=& \left(\frac{M}{E}\right)^2
\frac{E}{(k^0-p^0)^2 - (\bbox{k}_{\text{ET}}-\bbox{p}_{\text{ET}})^2 - \mu^2}\\
&=& \left(\frac{M}{E}\right)^2
\frac{E}{(k_{\text{ET}}-p_{\text{ET}})^2 - \mu^2},
\label{retpot}
\end{eqnarray}
where $k_{\text{ET}}$ and  $p_{\text{ET}}$ represent four-vectors,
defined by the equal-time three-vectors and the condition that
$k_{\text{ET}}^2=p_{\text{ET}}^2=M^2$.  These potentials resemble the
three-dimensional Blankenbecler-Sugar \cite{Blankenbecler:1966gx} or
Gross \cite{Gross:1969rv} quasi-potentials.

Both of these approximations are reasonable if the energy difference
between the initial and final states is small, which is valid for
lightly-bound states.  The instantaneous
potential is a better approximation of the symmetrized-mass potential,
since if we expand the symmetrized-mass potential to second-order in
perturbation theory about $k^0=p^0$, we get $V_\omega=V_{\text{Inst}}$.
Also, note that these potentials are explicitly rotationally invariant
in terms of our equal-time parameterization, which provides significant
computational advantages.

\subsection{Three-dimensional reduction of the Bethe-Salpeter equation}
\label{sec:3dredbasic}

We now consider a non-perturbative approximation used by Wallace and
Mandelzweig \cite{wallace,Wallace:1989nm}.  The basic idea is to first
make an approximation of the Bethe-Salpeter equation, then reduce that
modified Bethe-Salpeter equation to the physically 
equivalent Hamiltonian equation.
This approach was used by Phillips and Wallace \cite{Phillips:1996eb}
for the model we use, however, they obtained an equal-time Hamiltonian,
while we seek a light-front Hamiltonian.  Before we do this,  we first
review the basic mechanics of the three-dimensional reduction, as
presented in Sales {\it et al.} \cite{Sales:1999ec} and specialized to
our particular case.  We postpone the discussion of the approximation
until section \ref{sect:3dred}.

The Bethe-Salpeter equation can be written in matrix form as
$\Gamma=KG_0\Gamma$, or explicitly in function form in the momentum
basis as
\begin{eqnarray}
\Gamma(k_1;P) &=& \int \frac{d^4 p_1}{(2\pi)^4} K(k_1,p_1;P) G_0(p_1;P)
\Gamma(p_1;P).
\end{eqnarray}
In these equations, $\Gamma$ is the four-dimensional vertex function,
$K$ is the four-dimensional kernel, and $G_0$ is the two-particle
four-dimensional Green's function.  The momenta are $P$, the
total four-momentum, and $p_1$, the four-momentum of particle
1.  Particle 2's momentum is implicitly $P-p_1$. The four-dimensional 
Green's function is given by
\begin{eqnarray}
G_0(k_1;P) &=& i d(k_1) d(P-k_1),
\end{eqnarray}
where $d$ is the one-particle Green's function.  On the light front,
$d$ can be written as
\begin{eqnarray}
d(p) &=& \left(\frac{1}{p^+}\right)
\frac{1}{p^- - \mbox{Sign}(p^+)\epsilon^-(\bbox{p})},
\label{onepartprop}
\end{eqnarray}
where the light-front energy $\epsilon^-$ is given by 
\begin{eqnarray}
\epsilon^-(\bbox{p}) &=& \frac{M^2 + \bbox{p}_\perp^2}{|p^+|} - i \eta.
\end{eqnarray}
The real part of $\epsilon^-$ is a positive definite quantity, and
$\eta$ is positive infinitesimal. 

The Bethe-Salpeter equation can be rewritten \cite{Woloshyn:1973} as
\begin{eqnarray}
\Gamma &=& W \widehat{G}_0 \Gamma, \label{modbse}
\end{eqnarray}
where $\widehat{G}_0$ is an auxiliary Green's function, and $W$ is
defined by
\begin{eqnarray}
W &=& K + K(G_0-\widehat{G}_0)W.
\end{eqnarray}
The advantage of this rearrangement is that we are free to choose the
form of the auxiliary Green's function, $\widehat{G}_0$.  The choice of
$\widehat{G}_0$ advocated in Ref.~\cite{Sales:1999ec} is
\begin{eqnarray}
\widehat{G}_0(k_1,p_1;P)
&=& 
G_0(k_1;P)
\frac{\delta^{(2,+)}(\bbox{k}_1-\bbox{p}_1)}{g_0(\bbox{k}_1;P)}
G_0(p_1;P), \label{defghat}
\end{eqnarray}
where
\begin{eqnarray}
g_0(\bbox{k}_1,P) &=& \int \frac{dk_1^-}{2(2\pi)}
G_0(k_1;P). \label{defg0}
\end{eqnarray}
There is an extra factor of 2 in the denominator of Eq.~(\ref{defg0})
when compared to the equal-time formalism.  This is due to the Jacobian
of the light-front coordinates.

Using the definition of $\widehat{G}_0$ given in Eq.~(\ref{defghat}),
we can integrate the modified Bethe-Salpeter equation, in
Eq.~(\ref{modbse}), over the light-front energy to get
\begin{eqnarray}
\gamma(\bbox{k}_1;P) &=& \int \frac{d^2p_{1,\perp} \, dp^+_1}{(2\pi)^3}
w(\bbox{k}_1,\bbox{p}_1;P) g_0(\bbox{p}_1;P) \gamma(\bbox{p}_1;P),
\label{eqn:3ered}
\end{eqnarray}
where
\begin{eqnarray}
w(\bbox{k}_1,\bbox{p}_1;P) &\equiv& 
\frac{1}{g_0(\bbox{k}_1;P)}
\langle G_0 W G_0 \rangle (\bbox{k}_1,\bbox{p}_1;P)
\frac{1}{g_0(\bbox{p}_1;P)} \\
\gamma(\bbox{k}_1;P) &\equiv & \frac{1}{g_0(\bbox{k}_1;P)}
\int \frac{dk^-_1}{2(2\pi)} G_0(k_1;P) \Gamma(k_1;P).
\end{eqnarray}
The functional $\langle f \rangle$ is defined by its action on an
arbitrary function $f(k_1,p_1)$, where $k_1$ and $p_1$ are four-vectors,
as
\begin{eqnarray}
\langle f \rangle (\bbox{k}_1,\bbox{p}_1) &=& \int
\frac{dk^-_1}{2(2\pi)} \, \frac{dp^-_1}{2(2\pi)} f(k_1,p_1).
\end{eqnarray}

We proceed by calculating the specific form of $g_0$,
\begin{eqnarray}
g_0(\bbox{k}_1;P)
&=& \frac{\theta(k_1^+)\theta(k_2^+)}{2 k_1^+ k_2^+}
\frac{1}{P^- -k_1^- -k_2^-},
\end{eqnarray}
where $k_i^-=\epsilon^-(\bbox{k}_i)$.  As $g_0$ is a three-dimensional
quantity, it is clear that $k_i^-$ is not the independent minus
component of a momentum four-vector.  With this expression for $g_0$, we
can specialize Eq.~(\ref{eqn:3ered}) to the center-of-momentum frame and
obtain 
\begin{eqnarray}
\left( E -k_1^- -k_2^- \right) \psi(\bbox{k}_1;E)
&=& \int d^2p_{1,\perp} \int_0^{E} dp_1^+ \,
\frac{w(\bbox{k}_1,\bbox{p}_1;E)}{2(2\pi)^3 \sqrt{k_1^+ k_2^+ p_1^+ p_2^+}}
\psi(\bbox{p}_1;E),
\end{eqnarray}
where
\begin{eqnarray}
\psi(\bbox{k}_1;E) &=&
\frac{g_0(\bbox{k}_1;E)}{\sqrt{k_1^+k_2^+}} 
\gamma(\bbox{k}_1;E).
\end{eqnarray}
By comparing this equation to Eq.~(\ref{fullse3}), we find that the 
full light-front two-nucleon effective potential that corresponds to
the kernel $K$, after suppressing the coefficient given in
Eq.~(\ref{supressfactor}), is 
\begin{eqnarray}
V(\bbox{k}_1,\bbox{p}_1;E) &=& \frac{1}{E}
w(\bbox{k}_1,\bbox{p}_1;E). \label{convredtopot}
\end{eqnarray}
Thus, we can calculate light-front potentials directly from the
Bethe-Salpeter equation using this method.

The potential $V$ can be expanded in powers of the coupling constant, as
done in LFTOPT.  We find that when the auxiliary Green's function given
in Eq.~(\ref{defghat}) and the OBE kernel are used, the lowest order
parts of the potential (as calculated in \cite{Sales:1999ec}) are the
same as our OBE and TBE:SB potentials.
Thus, we conclude that this method
produces the physically 
equivalent Hamiltonian theory to the Bethe-Salpeter equation
being used.  We will use this in the next section to derive a
Hamiltonian potential for a situation where LFTOPT cannot be used.

\subsection{The modified-Green's-function approach}
\label{sect:3dred}

Now that the technology for the three-dimensional reduction has been
reviewed, we derive an approximate kernel for the Bethe-Salpeter
equation.  We will follow the approach of Phillips and Wallace
\cite{Phillips:1996eb} and works cited therein.  The idea is to start
with the Bethe-Salpeter equation where the kernel is truncated to only
include ladder (one-boson-exchange) and crossed (two-boson-exchange)
parts,
\begin{eqnarray}
\Gamma &=& (K_{\text{ladder}}+K_{\text{cross}}) G_0 \Gamma.
\label{laddercrossBSE}
\end{eqnarray}
An uncrossed approximation is used where the crossed part of the kernel
is approximated by
$K_{\text{cross}}\approx K_{\text{ladder}}G_CK_{\text{ladder}}$.
Our job is to find a valid modified Green's function, $G_C$.  Using this
uncrossed approximation,
\begin{eqnarray}
\Gamma &\approx& (K_{\text{ladder}}+
K_{\text{ladder}} G_C K_{\text{ladder}}) G_0 \Gamma.
\label{ladderuncrossBSEstart}
\end{eqnarray}

One can attempt to rewrite Eq.~(\ref{laddercrossBSE}) as an equation
linear in $K_{\text{ladder}}$, to obtain the modified-Green's-function
Bethe-Salpeter equation,
\begin{eqnarray}
\Gamma_{\text{MGF}} &=& K_{\text{ladder}} \Big( G_0 + G_C \Big) 
\Gamma_{\text{MGF}}.
\label{ladderuncrossBSE}
\end{eqnarray}
By iterating this integral equation for $\Gamma_C$, we obtain
\begin{eqnarray}
\Gamma_{\text{MGF}} &=&
\left[
K_{\text{ladder}} + \sum_{n=1}^\infty
K_{\text{ladder}} \left(G_C K_{\text{ladder}}\right)^n \right] 
G_0 \Gamma_{\text{MGF}}.
\end{eqnarray}
The part of Eq.~(\ref{ladderuncrossBSE}) that plays the role of the
kernel includes the uncrossed approximation of the original kernel
$K_{\text{ladder}}+K_{\text{ladder}}G_C K_{\text{ladder}}$ as well as
many more terms.  We note that the higher-order terms approximate some
of the higher-order terms that should be included in the full kernel,
such as three-boson-exchange diagrams where several meson lines cross.
However, this approach undercounts the higher-order
terms which it approximates, and also leaves out some terms completely.
Therefore, this new Bethe-Salpeter equation will give results that are
closer to the full solution than Eq.~(\ref{laddercrossBSE}), but will
not give the exact solution.  The articles by Wallace and
Mandelzweig \cite{wallace,Wallace:1989nm} demonstrate that this
approach, by effectively summing an infinite set of interactions, gives
the correct one-body limit, which is something that the usual
Bethe-Salpeter equation with a truncated kernel cannot do.

The modified Bethe-Salpeter equation in Eq.~(\ref{ladderuncrossBSE}) is
reduced to a Hamiltonian equation via the technique discussed in the
previous section.  
The equal-time Hamiltonian has been derived by Phillips and Wallace
\cite{Phillips:1996eb}.   They found that this modified-Green's-function
approach gave a spectra that lies closer to the full ground-state
spectra than the other approximations they considered.  We will use the
light-front reduction to obtain the light-front potential for the
Hamiltonian equation physically equivalent to Eq.~(\ref{ladderuncrossBSE}).

\begin{figure}[!ht]
\begin{center}
\epsfig{angle=0,width=4.112in,height=1.573in,file=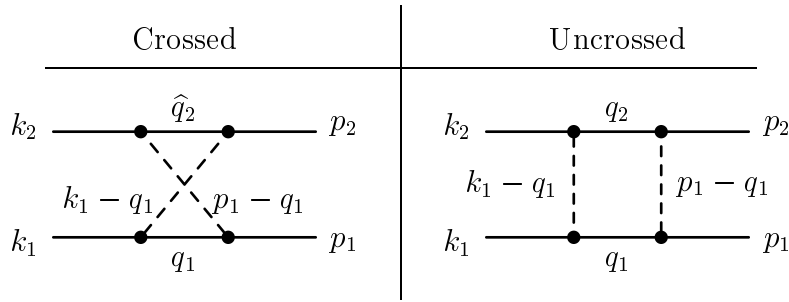}
\caption{The crossed and uncrossed Feynman graphs, where
$\widehat{q}_2=-q_2+p_2+k_2$.
\label{crosseduncrosseddiagrams}} 
\end{center}
\end{figure}

To clearly see what role $G_C$ plays, we compare the crossed and
uncrossed Feynman graphs in Fig.~\ref{crosseduncrosseddiagrams}.
Using the Feynman rules,
\begin{eqnarray}
K_{\text{crossed}} &\propto& \int \frac{d^4q_1}{(2\pi)^4}
\frac{1}{(k_1-q_1)^2-\mu^2}
d(q_1) d(\widehat{q}_2) 
\frac{1}{(p_1-q_1)^2-\mu^2} \\
K_{\text{uncrossed}} &\propto& \int \frac{d^4q_1}{(2\pi)^4}
\frac{1}{(k_1-q_1)^2-\mu^2}
d(q_1) d(q_2)
\frac{1}{(p_1-q_1)^2-\mu^2},
\end{eqnarray}
where $d$ is the one-particle propagator given in
Eq.~(\ref{onepartprop}), and $\widehat{q}_2 = -q_2+p_2+k_2$.  The only
difference between these two graphs is that the crossed one has
$d(\widehat{q}_2)$ while the uncrossed one has $d(q_2)$.

We want an approximate one-particle Green's function
$d_C$ that only depends on $q_1$ and $P$, so that
\begin{eqnarray}
d_C(q_1;P) &\approx& d(\widehat{q}_2).
\end{eqnarray}
Substitution of $d_C(q_1;P)$ for $d(\widehat{q}_2)$ in the crossed graph
causes the graph to become uncrossed.  The penalty for this
simplification is that a modified Green's function
propagates in the intermediate state, namely $i d(q_1) d_C(q_1;P)$.  It
is important that this approximation is invariant under relabeling
particle labels, so we explicitly symmetrize by defining
\begin{eqnarray}
G_C &=& \frac{i}{2}\left[ d(q_1) d_C(q_1;P) + d(q_2) d_C(q_2;P) \right]
\end{eqnarray}

How should we approximate $d_C$?  Since we are interested in obtaining a
bound state, a low-energy approximation is chosen.  Specializing to the
center-of-momentum frame in this limit, the 
external momenta are half the total momentum, so $p_2 = k_2 = P/2$ and
$\widehat{q}_2 = P - q_2 = q_1$.  This approximation is similar to
the one used by Phillips and Wallace.  Thus, we define
$d_C(q_1;P) \equiv d(q_1)$ so
\begin{eqnarray}
G_C(k_1;P) &=& G_1(k_1;P) + G_2(k_1;P),
\end{eqnarray}
where we define $G_1$ and $G_2$ by 
\begin{eqnarray}
G_1(k_1;P) &=& \frac{i}{2} d(k_1)^2 \\
G_2(k_1;P) &=& \frac{i}{2} d(P-k_1)^2.
\end{eqnarray}
This approximation for $G_C$ is valid for this model for the
energy range we study, as discussed in Appendix \ref{uxcheck}.  

We can write the modified-Green's-function Bethe-Salpeter equation as
\begin{eqnarray}
\Gamma_{\text{MGF}} = K_{\text{ladder}} \widetilde{G}_0 \Gamma_{\text{MGF}}, \label{eqn:bsemod}
\end{eqnarray}
where
\begin{eqnarray}
\widetilde{G}_0 &=& G_0 + G_1 + G_2.
\end{eqnarray}
This is used as the starting point of three-dimensional reduction
discussed in Section \ref{sec:3dredbasic}, where $\widetilde{G}_0$ is
considered as the Green's function.  Before doing the reduction,
note that the two poles of $G_1$ and $G_2$ lie in the same half plane
for each function, so
\begin{eqnarray}
\widetilde{g}_0(\bbox{k}_1,P)
&\equiv& \int \frac{dk_1^-}{2(2\pi)} \widetilde{G}_0(k_1,;P) \\
&=& g_0(\bbox{k}_1,P)
\end{eqnarray}

Proceeding with the three-dimensional reduction of
Eq.~(\ref{eqn:bsemod}) in the center-of-momentum frame, we obtain
\begin{eqnarray}
\left( E -k_1^- -k_2^- \right) \widetilde{\psi}(\bbox{k}_1;E)
&=& \int d^2p_{1,\perp} \int_0^{E} dp_1^+ \,
\frac{\widetilde{w}(\bbox{k}_1,\bbox{p}_1;E)}
{2(2\pi)^3 \sqrt{k_1^+ k_2^+ p_1^+ p_2^+}}
\widetilde{\psi}(\bbox{p}_1;E),
\label{eqn:3eredtilde}
\end{eqnarray}
where
\begin{eqnarray}
\widetilde{w}(\bbox{k}_1,\bbox{p}_1;P) &=&
\frac{1}{g_0(\bbox{k}_1;P)}
\langle \widetilde{G}_0 \widetilde{W} \widetilde{G}_0 \rangle
(\bbox{k}_1,\bbox{p}_1;P)
\frac{1}{g_0(\bbox{p}_1;P)} \\
\widetilde{\psi}(\bbox{k}_1;E) &=&
\frac{1}{\sqrt{k_1^+k_2^+}} 
\int \frac{dk^-_1}{2(2\pi)} \widetilde{G}_0(k_1;P) \Gamma_{\text{MGF}}(k_1;P),
\end{eqnarray}
and the modified kernel $\widetilde{W}$ is given by
\begin{eqnarray}
\widetilde{W} &=&
K_{\text{ladder}} +
K_{\text{ladder}} 
\left[ \widetilde{G}_0 - \widehat{\widetilde{G}}_0 \right]
\widetilde{W} \\
\widehat{\widetilde{G}}_0(k_1,p_1;P)
&=& 
\widetilde{G}_0(k_1;P)
\frac{\delta^{(2,+)}(\bbox{k}_1-\bbox{p}_1)}{g_0(\bbox{k}_1;P)}
\widetilde{G}_0(p_1;P).
\end{eqnarray}
It is a feature of the light front that $\widetilde{g}_0=g_0$, so that
the uncrossed approximation only affects the potential, and
Eq.~(\ref{eqn:3eredtilde}) has the same form as Eq.~(\ref{eqn:3ered}).
In the equal-time calculation \cite{Phillips:1996eb}
$\widetilde{g}_0\neq g_0$, so the approximation changes both the Green's
function as well as the potential.

We now expand $\widetilde{W}$ in powers of the coupling constant, and
keep only the lowest order term, $K_{\text{ladder}}$.  According to
Eq.~(\ref{convredtopot}), the light-front potential that corresponds to
this truncation of the kernel is the modified-Green's-function (MGF)
potential $V_{\text{MGF}}$,
\begin{eqnarray}
V_{\text{MGF}}(\bbox{k}_1,\bbox{p}_1;P) &=& \frac{1}{E}
\frac{1}{g_0(\bbox{k}_1;P)}
\langle \widetilde{G}_0 K_{\text{ladder}} \widetilde{G}_0 \rangle
(\bbox{k}_1,\bbox{p}_1;P)
\frac{1}{g_0(\bbox{p}_1;P)}.
\end{eqnarray}
The one-boson-exchange kernel $K_{\text{ladder}}$ is given by the
Feynman diagram, so
\begin{eqnarray}
K_{\text{ladder}}(k_1,p_1;P) &=&
\frac{(iM)^2}{(k_1-p_1)^2-\mu^2+i\eta} \nonumber \\
&=&
\left( \frac{1}{k_1^+-p_1^+} \right)
\frac{M^2}{(k_1^--p_1^-) - \mbox{Sign}(k_1^+-p_1^+)
\omega^-(\bbox{k}_1-\bbox{p_1})},
\end{eqnarray}
where the light-front energy of the meson is given by
\begin{eqnarray}
\omega^-(\bbox{q}) &=& \frac{\mu^2 - {\bf q}_\perp^2}{|q^+|} - i \eta.
\end{eqnarray}

By examining the locations of all the poles in the $k^-$ integrals for 
$V_{\text{MGF}}$, we find the integrals are non-vanishing only when
both $x$ and $y$ are between $0$ and $1$.  The sign functions in the
denominator of $K_{\text{ladder}}$ naturally divide $V_{\text{MGF}}$
into two parts, one for $x<y$ and the other for $x>y$.  The integrals in
$V_{\text{MGF}}$ are straightforward, but quite lengthy and tedious.
Therefore, we show only the final answer,
\begin{eqnarray}
V_{\text{MGF}}(\bbox{k}_1,\bbox{p}_1;P)
&=& \left(\frac{M}{E}\right)^2 \left[
\frac{\theta(x-y)}{|x-y|} \left( \frac{1}{D_1} +
\frac{N_{p,21}+N_{k,12}}{2D_1^2} +
\frac{N_{p,21} N_{k,12}}{2D_1^3}
\right)
\right.\nonumber \\ & &
\phantom{
\left(\frac{M}{E}\right)^2 \left[
\right. } \left.+
\frac{\theta(y-x)}{|y-x|} \left( \frac{1}{D_2} +
\frac{N_{k,21}+N_{p,12}}{2D_2^2} + 
\frac{N_{p,12} N_{k,21}}{2D_2^3}
\right) \right], \label{mgfpot}
\end{eqnarray}
where
\begin{eqnarray}
N_{k,12} &=& \frac{k_1^+}{k_2^+}(E - k_1^- - k_2^-) \\
N_{k,21} &=& \frac{k_2^+}{k_1^+}(E - k_1^- - k_2^-) \\
D_1 &=& E - p_1^- - k_2^- - \omega^-(\bbox{k}_1-\bbox{p}_2) \\
D_2 &=& E - k_1^- - p_2^- - \omega^-(\bbox{p}_1-\bbox{k}_2).
\end{eqnarray}
The expressions for $N_{p,12}$ and $N_{p,21}$ are obtained by replacing
$k$ with $p$ in $N_{k,12}$ and $N_{k,21}$.

What is the physical interpretation of this modified-Green's-function
potential?  The first term multiplying each $\theta$ function gives the
OBE potential we derived before from the perturbation theory.  There the
$D$ in the denominators corresponds to one meson exchange.  The second
and third terms multiplying the $\theta$ functions, with $D^2$ and $D^3$
in the denominators appear to be effective two- and three-meson-exchange
terms.  Since time-ordered perturbation theory does not apply to the
modified Bethe-Salpeter equation that we use, the exact nature of these
terms is not easy to understand.  However, it is clear that these terms
increase the strength of the potential, and should mimic the
higher-order diagrams that are not being included explicitly.

The only dependence on the direction of the perpendicular components of
$k$ and $p$ comes from the $D$'s.  This allows the azimuthal-angle
integration of $V_{\text{MGF}}$ to be done easily, as shown in Appendix
\ref{angint}. 

\section{Results} \label{results}

For our numerical work, we pick the meson mass to be $0.15$ times that
of the nucleon, so $\mu = 0.15 M$.  This is chosen so that our ground
state can be considered a toy model of deuterium, and also to facilitate
comparison with the results of Nieuwenhuis and Tjon
\cite{Nieuwenhuis:1996mc} and Phillips and Afnan
\cite{Phillips:1996ed}.  Nieuwenhuis and Tjon used the Feynman-Schwinger
representation (FSR) of the two-particle Green's
function\cite{Simonov:1993kp} in the quenched approximation without the
mass and vertex renormalization terms \cite{Nieuwenhuis:1996mc}.  Their
result is to be considered the full solution that the Bethe-Salpeter and
Hamiltonian equations approximate.  For the Bethe-Salpeter equation,
computation of the bound-state energies for models similar to ours have
been done for the ladder \cite{schwartz} and ladder plus crossed
\cite{LevineWright3} kernels over 30 years ago.  More recent results are
found in  \cite{Phillips:1996ed,Theussl:1999xq}, where the solutions are
compared to those given by the FSR approach.

Now consider how the light-front Hamiltonian approach fits in with the
other approaches.  As discussed in section \ref{tbe:sbSec}, different
light-front potentials can be derived from Bethe-Salpeter equations with
different kernels.  We have mentioned that the OBE+TBE:SB potential
should approximate the ladder Bethe-Salpeter equation, and similarly the
OBE+TBE potential should approximate the Bethe-Salpeter equation when
the ladder plus crossed kernel is used.  The best that these truncated
Hamiltonians can do is approximate their respective Bethe-Salpeter
equations.

\begin{figure}[!ht]
\begin{center}
\epsfig{angle=0,width=5.0in,height=4.0in,file=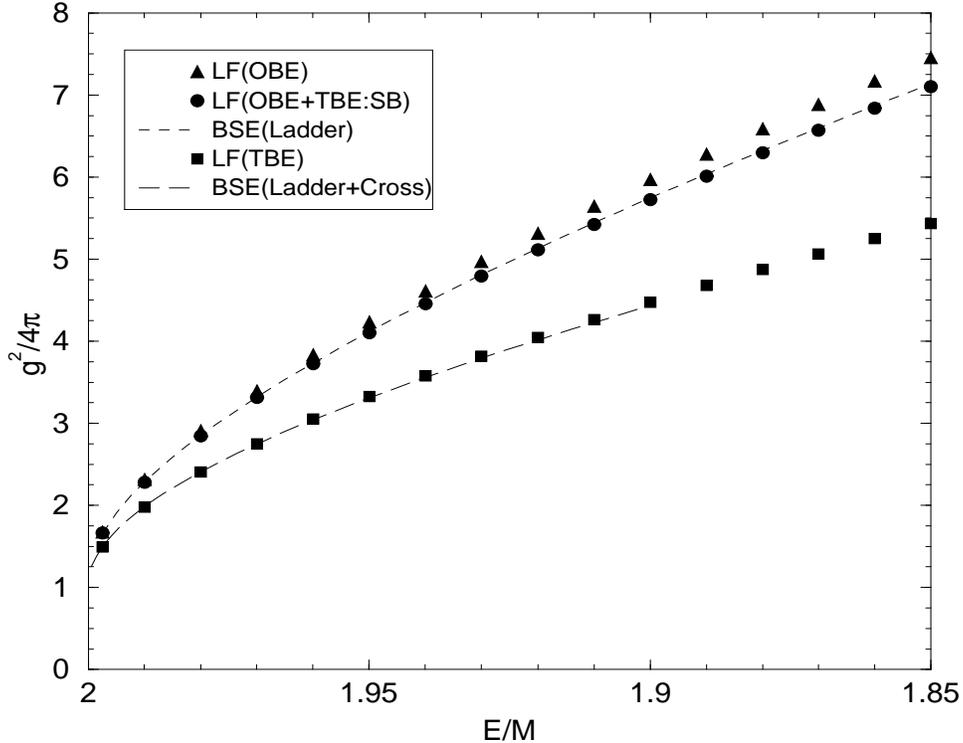}
\caption{We display here three light-front calculations using the OBE,
OBE+TBE:SB, and OBE+TBE potentials. The spectra for the Bethe-Salpeter
equation for the ladder and ladder plus crossed kernels are also
plotted.  The curves for the ladder 
(ladder plus crossed) Bethe-Salpeter equation and the light-front
OBE+TBE:SB (OBE+TBE) potentials are very close to each other, almost
indistinguishable in this figure.  $E$ is the energy of the ground state
of two nucleons, and $M$ is the mass of the nucleons.  The meson mass is
$\mu = 0.15 M$.  In terms of the binding energy $B$, $E=2M-B$.
\label{bslffig}}
\end{center}
\end{figure}

With this in mind, we evaluate the coupling constant versus bound-state
energy curves (which we will call the spectrum) for the Hamiltonian
equation with the OBE potential, the OBE+TBE:SB potential, and the
OBE+TBE potential.  For the range of values we use here, we find
numerical errors in the value of $g^2$ are less than 2\%.
Our results (without error bars) are plotted along with the results
obtained with the ladder BSE \cite{Phillips:1996ed}, and ladder plus
crossed BSE \cite{dandata} in Fig.~\ref{bslffig}.  We note that the
OBE+TBE:SB potential agrees well with the ladder BSE, and the OBE+TBE
potential agrees with the ladder plus crossed BSE.  This is the best
that a Hamiltonian can do, so this result is interpreted as evidence
that, in general, the higher-order diagrams are very small for the
ground state on the light front.

\begin{figure}[!ht]
\begin{center}
\epsfig{angle=0,width=5.0in,height=4.0in,file=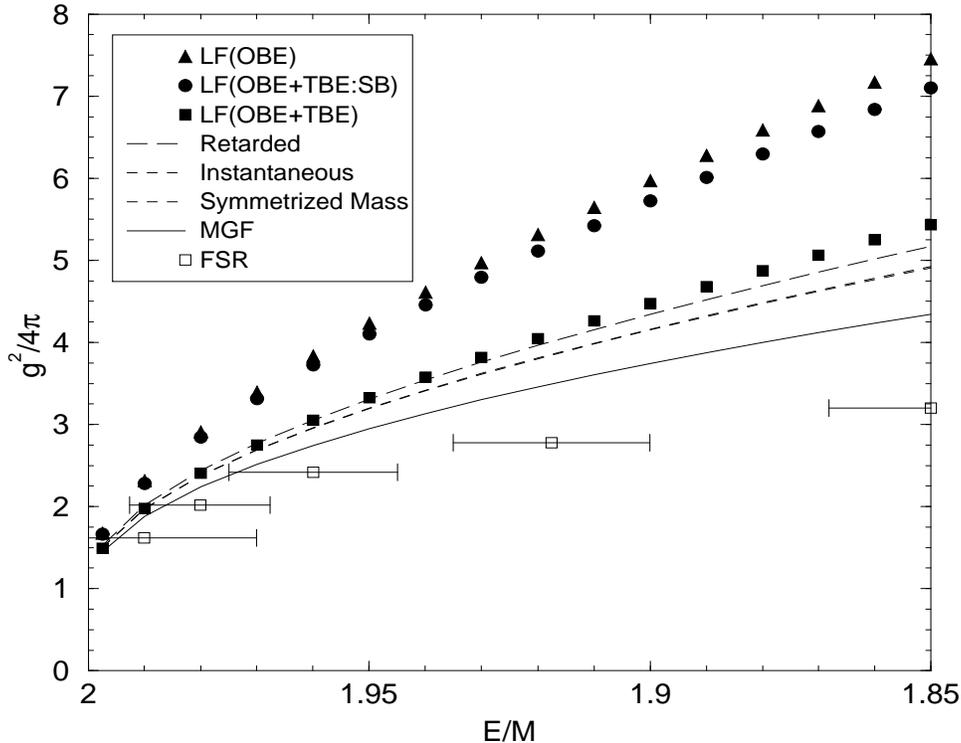}
\caption{The spectra for the seven light-front
potentials (OBE, OBE+TBE:SB, OBE+TBE, retarded, instantaneous,
symmetrized mass, modified-Green's-function) are plotted here along
with the spectra for the full solution (FSR).  The instantaneous curve
lies very close to the symmetrized mass curve, hence the same line style
is used for both.
\label{sumfig}}
\end{center}
\end{figure}

If all one wanted was a way to approximate the spectra for
Bethe-Salpeter equations with different kernels, one could just use the
truncated potentials that the LFTOPT provide.  However, the true goal is
to approximate the spectra for the full ground state, which in this
model is given by the FSR approach \cite{Nieuwenhuis:1996mc}.  We expect
that the non-perturbative potentials should give a better approximation
of the full solution than the perturbative potentials, since the
non-perturbative potentials attempt to incorporate physics from
higher-order diagrams, although this is not immediately clear by looking
at the forms of the potentials used. We plot the results for all of the
light-front potentials described in this paper, the three truncated
potentials (OBE, OBE+TBE:SB, and OBE+TBE) and the four non-perturbative
potentials (symmetrized mass, retarded, instantaneous, and modified
Green's function) along with the results for the full theory in
Fig.~\ref{sumfig}.  For deeply-bound states, there is considerable
disagreement between the perturbative results and the full results,
while the non-perturbative results do better, with the
modified-Green's-function (MGF) potential achieving the closest
agreement.  For lightly-bound states, the results for all of the
potentials appear converge to each other, close to the full result.

For the modified-Green's-function potential, only the first term of the
expansion in $g^2$ was kept.  In principle, higher-order terms could
be calculated.  However, since there was fairly good agreement between
the OBE potential and the ladder Bethe-Salpeter equation, it is expected
that the MGF potential will give results that are close to the
ladder Bethe-Salpeter equation using the modified-Green's-function.

\section{Conclusions} \label{conclusions}

In this paper, we consider the ground state of a massive Wick-Cutkosky
model using a Hamiltonian derived from light-front dynamics.  We examine
three different truncations of the effective potential derived from the 
perturbative field theory, and four approximations that attempt to
incorporate non-perturbative physics.  For each of these potentials, we
calculate the coupling constant that gives the ground-state for a given
bound-state energy (the spectra), and compare to the spectra from
different approaches found in the literature.  We find fairly good
agreement between all the methods for lightly-bound systems.

For the full range of binding energies studied, the results for
calculations including one- and two-boson-exchange potentials agree with
the Bethe-Salpeter equation using the physically equivalent truncation
of the kernel  within the numerical errors.  (This is a consequence of
examining the ground state.  For the excited states more higher-order
light-front time-ordered graphs are required to get the same level of
agreement \cite{Cooke:1999yi}.)  The agreement for the case with the
stretched box diagrams (OBE+TBE:SB) has been shown previously by Sales
{\it et al.} \cite{Sales:1999ec}; the result with the crossed-box
contribution (OBE+TBE) is new.  This excellent agreement with the
Bethe-Salpeter results has an undesirable consequence: The BSE results
are known to be a poor approximation of the full solution for deeply
bound systems, so the truncated Hamiltonian approach cannot provide a
good approximation to the full solution in that regime.

The non-perturbative potentials based on physical considerations give a
better approximation of the full solution than the potentials obtained
from LFTOPT.  For all binding energies, the modified-Green's-function
potential achieves the closest agreement with the full solution of
all the potentials considered here.  However, there is still
considerable disagreement between approximate potentials and the full
result for deeply bound states.  We interpret this as an indication that
the approximations, while incorporating some non-perturbative physics,
do not go far enough.  In the weakly-bound regime, which is of
relevance for deuteron calculations, the spectra for all of the
potentials are close together, indicating that light-front dynamics
provides a good description of lightly-bound systems.

\acknowledgements

We are grateful to D.R.~Phillips for extensive
discussions and for providing us with unpublished material.
This work is supported in part by the U.S.~Dept.~of Energy under Grant
No.~DE-FG03-97ER4014.  
\appendix

\section{Notation, conventions, and useful relations}
\label{notation}

This is patterned after the review by
Harindranath\cite{Harindranath:1996hq}. For a general four-vector $a$,
we define the light-front variables 
\begin{eqnarray}
a^\pm          &=& a^0 \pm a^3, \\
\bbox{a}_\perp &=& (a^1,a^2),
\end{eqnarray}
so the 4-vector $a^\mu$ can be denoted
\begin{eqnarray}
a&=&(a^+,a^-,\bbox{a}_\perp).
\end{eqnarray}
Using this, we find that the scalar product is
\begin{eqnarray}
a \cdot b &=& a^\mu b_\mu
= \frac{1}{2}\left(a^+ b^- + a^- b^+ \right) - \bbox{a}_\perp \cdot 
\bbox{b}_\perp.
\end{eqnarray}
This defines $g_{\mu\nu}$, with $g_{+-}=g_{-+}=1/2$, $g_{11}=g_{22}=-1$,
and all other elements of $g$ vanish.  The elements of $g^{\mu\nu}$ are
obtained from the condition that $g^{\mu\nu}$ is the inverse of
$g_{\mu\nu}$, so
$g^{\alpha\beta}g_{\beta\lambda}=\delta^\alpha_\lambda$.  Its elements
are the same as those of $g_{\mu\nu}$, except for
$g^{-+}=g^{+-}=2$. Thus,
\begin{eqnarray}
a^\pm &=& 2 a_\mp.
\end{eqnarray}
and the partial derivatives are similarly given by
\begin{eqnarray}
\partial^\pm &=& 2 \partial_\mp = 2 \frac{\partial}{\partial x^\mp}.
\end{eqnarray}

Moving to the physical consequences of this coordinate system, 
the commutation relations $[p^\mu,x^\nu] = i g^{\mu\nu}$ yields 
\begin{eqnarray}
\,[       p^\pm     ,      x^\mp      ] &=& 2i \\
\,[ \bbox{p}_\perp^i,\bbox{x}_\perp^j ] &=& -i \delta_{i,j},
\end{eqnarray}
with the other commutators equal to zero.  Thus, $\bbox{x}_\perp^i$ is
canonically conjugate to $\bbox{p}_\perp^i$, and $x^\pm$ is conjugate to
$p^\mp$.  In light-front dynamics, $x^+$ plays the role
of time (the light-front time), so $p^-$ is the light-front
energy and the light-front Hamiltonian is given by $P^-$.

Particles have the light-front energy defined by the on-shell constraint
$k^2=m^2$.  This implies that the light-front energy is 
\begin{eqnarray}
k^- &=& \frac{m^2 +\bbox{k}_\perp^2}{k^+}.
\end{eqnarray}
The free components of the momentum can be written as the light-front
three-vector $\bbox{k}$, denoted by
\begin{eqnarray}
\bbox{k} &=& (k^+,\bbox{k}_\perp).
\end{eqnarray}

\section{Conversion to matrix form}
\label{numericaleq}

To solve for the bound-state wavefunction numerically, the light-front
Schr\"{o}dinger equation given in Eq.~(\ref{fullse6}) must be
discretized and cast in matrix form.  The equation is first 
symmetrized to get
\begin{eqnarray}
-\int_0^\infty dp_{\text{ET}} \int_0^{\pi/2} d\theta_p
V_S^+(k_{\text{ET}},\theta_k;p_{\text{ET}},\theta_p)
\psi_S^{\text{GS}}(p_{\text{ET}},\theta_p)
&=&
\psi_S^{\text{GS}}(k_{\text{ET}},\theta_k),
\label{fullse7}
\end{eqnarray}
where
\begin{eqnarray}
V_S^+(k_{\text{ET}},\theta_k;p_{\text{ET}},\theta_p)
&=&
B(k_{\text{ET}})
A(k_{\text{ET}},\theta_k)
V^+(k_{\text{ET}},\theta_k;p_{\text{ET}},\theta_p)
A(p_{\text{ET}},\theta_p)
B(p_{\text{ET}})
\\
\psi_S^{\text{GS}}(k_{\text{ET}},\theta_k)
&=&
B(k_{\text{ET}})^{-1} A(k_{\text{ET}},\theta_k)
\psi^{\text{GS}}(k_{\text{ET}},\theta_k) \\
A(k_{\text{ET}},\theta_k) &=&
\sqrt{\frac{2 k_1^+ k_2^+  k^2_{\text{ET}} \sin\theta_k}{(k^0)}} \\
B(k_{\text{ET}}) &=& \sqrt{\frac{-1}{E^2 - 4 (k^0)^2}}.
\end{eqnarray}

Before discretizing the integrals, note that
\begin{eqnarray}
\int_0^\infty dp \, f(p) &=& 
a \int_0^1 du \left(f(au) + \frac{f(a/u)}{u^2} \right). \label{inttrick}
\end{eqnarray}
Using this trick, the $p_{\text{ET}}$ integral in Eq.~\ref{fullse7} can
be written as an integral over a finite range.  Since we are concerned
with a bound state, the wavefunction is exponentially damped for large
momenta, and the second term of Eq.~(\ref{inttrick}) converges as $u$
approaches zero.

All the integrals in Eq.~\ref{fullse7} then are over a
finite range, and can be discretized using Gauss-Legendre quadrature.
The specific routines for the quadrature are
given by {\it Numerical Recipes in C} \cite{press}.  This conversion gives a
matrix equation that approximates the original Eq.~\ref{fullse7},
\begin{eqnarray}
-V_S^+(g(E),E) \psi_S^{\text{GS}} &=& \psi_S^{\text{GS}},
\label{fullse8}
\end{eqnarray}
where the explicit dependence of $V_S^+$ on the symmetrized potential on
the binding energy $E$ and the coupling constant $g$ is shown.  This
equation must be solved self-consistently for the spectrum $g(E)$.

The approach we use is to first solve for the spectrum for the OBE
potential.  The eigenvalue equation
\begin{eqnarray}
V_{S,\text{OBE}}^+(E)
\psi_S^{\text{GS}}
&=& \alpha
\psi_S^{\text{GS}},
\end{eqnarray}
where
\begin{eqnarray}
\alpha &=& \frac{-1}{g_{\text{OBE}}(E)^2}.
\end{eqnarray}
The ground-state wavefunction is the eigenvector that corresponds to the
smallest eigenvalue $\alpha$.  We calculate the wavefunction and the
smallest coupling constant using EISPACK \cite{eispack} routines for a
range of energies to map out the spectrum.

Using the coupling constant for the OBE potential as a starting point,
we can use Eq.~\ref{fullse8} for higher-order potentials that include
$N$ meson exchanges.  For a
given energy, the coupling constant $g(E)$ is initially chosen as
$g_{\text{OBE}}(E)$, then we solve
\begin{eqnarray}
\left[
\sum_{n=1}^N g(E)^{2n} V_{S,(2n)}^+(E)
\right]
\psi_S^{\text{GS}}
&=& \beta
\psi_S^{\text{GS}},
\end{eqnarray}
as an eigenvalue equation for $\beta$.  The coupling constant $g(E)$ is
varied until the the lowest eigenvalue is $\beta=-1$, at which point
$g(E)$ is the correct value of the spectrum corresponding to the
ground-state wavefunction $\psi_S^{\text{GS}}$.

\section{Azimuthal-angle integration of the OBE and MGF potentials}
\label{angint}

In this section, we evaluate the azimuthal-angle integration of the OBE
potential in Eq.~(\ref{OBEpot2}) and the first term in MGF potential in
Eq.~(\ref{mgfpot}), using the prescription for azimuthal-angle
integration given in Eq.~(\ref{angintlf}).  One of the integrals is
easily done since since the potential is independent of the azimuthal
angle between the two perpendicular momenta, so
\begin{eqnarray}
V(k^+,k_\perp;p^+,p_\perp) &\propto&
\left[
 \theta(x-y) \int_0^{2\pi} \frac{d\phi}{A_1 + B \cos\phi}
+\theta(y-x) \int_0^{2\pi} \frac{d\phi}{A_2 + B \cos\phi}
\right], \label{obeangint}
\end{eqnarray}
where
\begin{eqnarray}
A_1 &=& (k_1^+-p_1^+)(E-p^-_1-k^-_2)-\mu^2-p_\perp^2-k_\perp^2 \\
A_2 &=& (p_1^+-k_1^+)(E-k^-_1-p^-_2)-\mu^2-p_\perp^2-k_\perp^2 \\
B   &=& 2k_\perp p_\perp.
\end{eqnarray}
The integrals in Eq.~\ref{obeangint} are easily done to give, since the
$A$'s are negative,
\begin{eqnarray}
\int_0^{2\pi} \frac{d\phi}{A + B \cos\phi} &=&
\frac{-2\pi}{\sqrt{A^2-B^2}}. \label{int1}
\end{eqnarray}
Using this, the azimuthal-angle-averaged OBE potential is given by
\begin{eqnarray}
V_{\text{OBE}}(k^+,k_\perp;p^+,p_\perp) &=& -2\pi
\left(\frac{M}{E}\right)^2 E \left[
 \frac{\theta(x-y)}{\sqrt{A_1^2-B^2}}
+\frac{\theta(y-x)}{\sqrt{A_2^2-B^2}}
\right].
\end{eqnarray}
It is straightforward to rewrite this equation for the potential in terms
of the equal-time coordinates.

When the other terms in the MGF potential are azimuthal-angle averaged, 
integrations similar to the one given in Eq.~\ref{int1} are encountered,
with the denominator squared or cubed.  We note that
\begin{eqnarray}
\int_0^{2\pi} \frac{d\phi}{(A + B \cos\phi)^2} &=&
\frac{-2\pi}{\sqrt{A^2-B^2}} \, \frac{A}{A^2-B^2} \\
\int_0^{2\pi} \frac{d\phi}{(A + B \cos\phi)^3} &=&
\frac{-2\pi}{\sqrt{A^2-B^2}} \, \frac{2A^2+B^2}{(A^2-B^2)^2},
\end{eqnarray}
so the azimuthal-angle-averaged MGF potential is given by
\begin{eqnarray}
V_{\text{MGF}}(\bbox{k}_1,\bbox{p}_1;P)
&=& -2\pi \left(\frac{M}{E}\right)^2 \left[
\frac{\theta(x-y)}{\sqrt{A_1^2-B^2}} \left( 1 +
\frac{N_{p,21}+N_{k,12}}{2D_{1,2}} +
\frac{N_{p,21} N_{k,12}}{2D_{1,3}}
\right)
\right.\nonumber \\ & &
\phantom{
 -2\pi \left(\frac{M}{E}\right)^2 \left[
\right. } \left.+
\frac{\theta(y-x)}{\sqrt{A_2^2-B^2}} \left( 1 +
\frac{N_{k,21}+N_{p,12}}{2D_{2,2}} + 
\frac{N_{p,12} N_{k,21}}{2D_{2,3}}
\right) \right],
\end{eqnarray}
where
\begin{eqnarray}
D_{i,2} &\equiv& \frac{A^2-B^2}{A} \\
D_{i,3} &\equiv& \frac{(A^2-B^2)^2}{2A^2+B^2},
\end{eqnarray}
and $i=1,2$.

\section{Azimuthal-Angle Integration and Loop integration of the TBE
potentials} 
\label{loopints}

As in the previous section, we want the azimuthal-angle integrals of
the TBE potentials given in Eqs.~(\ref{TBESBpot}-\ref{TBEZXpot}).  For
these potentials, there is also a loop integral that has to be done.
We start by analyzing the equations schematically.  Each of the terms in
the TBE potentials can be written in the following form,
\begin{eqnarray}
V_{\text{TBE}}(k^+,k_\perp;p^+,p_\perp) &=&
\int_0^\infty \frac{q_\perp dq_\perp}{2(2\pi)^3} \int_0^1 dz \,
J(k^+,q^+,p^+)
I(k^+,k_\perp,q^+,q_\perp,p^+,p_\perp) \\
I(k^+,k_\perp,q^+,q_\perp,p^+,p_\perp) &=& 
\int_0^{2\pi} d\phi_q d\phi_p
\frac{1}{A_1+B_1\cos\phi_q}
\nonumber \\ & & \times
\frac{1}{A_2+B_2\cos\phi_q+C_2\cos\phi_p+D_2\cos(\phi_p-\phi_q)}
\nonumber \\ & & \times
\frac{1}{A_3+B_3\cos\phi_q+C_3\cos\phi_p+D_3\cos(\phi_p-\phi_q)},
\end{eqnarray}
where the $A$'s, $B$'s, $C$'s, and $D$'s may have dependence on $k^+$,
$k_\perp$, $p^+$, $p_\perp$, $q^+=zE$, and $q_\perp$; they are
independent of the azimuthal angles.  These functions can be easily
determined for  each potential by examining the forms of the original
equations.  The rotational invariance of the potential about the
three-axis allows the $\phi_k$ integration to be done trivially.

In the integrand of $I$, only the last two terms depend on $\phi_p$.  To
emphasize this, we write
\begin{eqnarray}
I(k^+,k_\perp,q^+,q_\perp,p^+,p_\perp) &=& 
\int_0^{2\pi} d\phi_q
\frac{I_2(k^+,k_\perp,\bbox{q},p^+,p_\perp)}
{(A_1+B_1\cos\phi_q)(A_2+B_2\cos\phi_q)(A_3+B_3\cos\phi_q)}
\\
I_2(k^+,k_\perp,\bbox{q},p^+,p_\perp) &=& 
\int_0^{2\pi} 
\frac{d\phi_p}{(1+a_2\cos\phi_p+b_2\sin\phi_p)(1+a_3\cos\phi_p+b_3\sin\phi_p)}
\end{eqnarray}
where, for $i=2,3$,
\begin{eqnarray}
a_i &=& \frac{C_i+D_i \cos\phi_q}{A_i+B_i \cos\phi_q} \\
b_i &=& \frac{    D_i \sin\phi_q}{A_i+B_i \cos\phi_q}.
\end{eqnarray}
The integral in $I_2$ is evaluated to obtain
\begin{eqnarray}
I_2(k^+,k_\perp,\bbox{q},p^+,p_\perp) &=& 
\frac{2\pi}{(a_2-a_3)^2 + (b_2-b_3)^2 - (a_2 b_3 - a_3 b_2)^2} \nonumber
\\ & & \times
\left(
\frac{a_2(a_2-a_3)+b_2(b_2-b_3)}{\sqrt{1-a_2^2-b_2^2}}
+
\frac{a_3(a_3-a_2)+b_3(b_3-b_2)}{\sqrt{1-a_3^2-b_3^2}}
\right)
\end{eqnarray}

The remaining three-dimensional loop integral in $V_{\text{TBE}}$ on
$\bbox{q}$ is done using numeric techniques.  The trick introduced in
Appendix \ref{numericaleq} to convert the semi-infinite $q_\perp$
integration into an integration on a compact range.  Before doing the
$z$ integral, the range of integration is limited by using the step
functions.  Gauss-Legendre quadrature, given by {\it Numerical Recipes
in C} \cite{press},  is used to evaluate all the integrals. 

Since each of the parts of the full TBE potential (TBE:SB, TBE:SX,
\ldots) should be hermitian and invariant under interchange of particle
1 and 2, these invariances can be used as a self-consistency check.
Each matrix element is calculated twice, first by using the
straightforward approach, then particle labels 1 and 2 are interchanged
and it is calculated again.  The results are compared, and if they
differ by an unacceptable amount, the number of quadrature points is
increased and the element is recalculated.  In order to get the
numerical accuracy of the potentials correct to within 1\%, 
we start with ten points for the $q_\perp$ integral, six points for the
$\phi_q$ integral, and three points for the $z$ integral, resulting in a
three-dimensional integral using 180 points.

\section{Check of the Uncrossed Approximation}
\label{uxcheck}

In this section, we want to check that how well the approximation
\begin{eqnarray}
K_{\text{cross}} &\approx&  K_{\text{ladder}} G_C K_{\text{ladder}},
\end{eqnarray}
works.  Since we are using a Hamiltonian theory and are interested in the
potentials, we compare the potentials defined by
\begin{eqnarray}
V_{\text{TBE:X}} &=& \frac{1}{E} g_0^{-1}
\langle G_0 K_{\text{cross}} G_0 \rangle g_0^{-1} \\
V_{\text{TBE:UX}} &=& \frac{1}{E} g_0^{-1}
\langle G_0 K_{\text{ladder}} G_C K_{\text{ladder}} G_0 \rangle g_0^{-1}.
\end{eqnarray}
The notation used here is defined in sections \ref{sec:3dredbasic} and
\ref{sect:3dred}.

The TBE crossed potential (TBE:X) can be written as
\begin{eqnarray}
V_{\text{TBE:X}} &=& 
V_{\text{TBE:SX}} +
V_{\text{TBE:TX}} +
V_{\text{TBE:WX}} +
V_{\text{TBE:ZX}},
\end{eqnarray}
where the potentials on the right-hand side are defined in section
\ref{deftbepots}.  Calculation of the TBE approximate uncrossed
potential (TBE:UX) is straightforward, but tedious.  We find that
\begin{eqnarray}
V_{\text{TBE:UX}}(E;\bbox{k}_1,\bbox{p}_1) &=&
\frac{1}{2} \Big[
V_{\text{TBE:UX1}}(E;\bbox{k}_1,\bbox{p}_1) +
V_{\text{TBE:UX1}}(E;\bbox{p}_1,\bbox{k}_1)
\nonumber \\ 
& &  \phantom{ \frac{1}{2} \Big[} +
V_{\text{TBE:UX2}}(E;\bbox{k}_1,\bbox{p}_1) +
V_{\text{TBE:UX2}}(E;\bbox{p}_1,\bbox{k}_1) \Big],
\end{eqnarray}
where
\begin{eqnarray}
V_{\text{TBE:UX1}}(E;\bbox{k}_1,\bbox{p}_1)
&=&
\left( \frac{M}{E} \right)^4
\int \frac{d^2 q_\perp}{2(2\pi)^3} \left[
\int_0^1 dz \frac{\theta(x-z) \theta(z-y)}{(x-z) z^2 (z-y)} 
\right. \nonumber \\ & & \phantom{
\left( \frac{M}{E} \right)^4
\int \frac{d^2 q_\perp}{2(2\pi)^3} \left[ \right.} \left. \times
\frac{1}{E - p_1^- - k_2^- - \omega^-(\bbox{q}_1-\bbox{p}_1)
- \omega^-(\bbox{k}_1-\bbox{q}_1)}
\right. \nonumber \\ & & \phantom{
\left( \frac{M}{E} \right)^4
\int \frac{d^2 q_\perp}{2(2\pi)^3} \left[ \right.} \left. \times
\left(
\frac{1}{E - q_1^- - k_2^- - \omega^-(\bbox{k}_1-\bbox{q}_1)}
\right)^2
 \right] \nonumber \\ & & \phantom{
\left( \frac{M}{E} \right)^4
\int \frac{d^2 q_\perp}{2(2\pi)^3}} 
+ \{ 1\leftrightarrow 2\} \\
V_{\text{TBE:UX2}}(E;\bbox{k}_1,\bbox{p}_1)
&=&
\left( \frac{M}{E} \right)^4
\int \frac{d^2 q_\perp}{2(2\pi)^3} \left[
\int_0^1 dz \frac{\theta(x-z) \theta(y-z)}{(x-z) z^2 (y-z)} 
\right. \nonumber \\ & & \phantom{
\left( \frac{M}{E} \right)^4
\int \frac{d^2 q_\perp}{2(2\pi)^3} \left[ \right.} \left. \times
\frac{1}{E - q_1^- - p_2^- - \omega^-(\bbox{p}_1-\bbox{q}_1)}
\right. \nonumber \\ & & \phantom{
\left( \frac{M}{E} \right)^4
\int \frac{d^2 q_\perp}{2(2\pi)^3} \left[ \right.} \left. \times
\left(
\frac{1}{E - q_1^- - k_2^- - \omega^-(\bbox{k}_1-\bbox{q}_1)}
\right)^2
 \right] \nonumber \\ & & \phantom{
\left( \frac{M}{E} \right)^4
\int \frac{d^2 q_\perp}{2(2\pi)^3}}
+ \{ 1\leftrightarrow 2\}.
\end{eqnarray}

Now consider the azimuthal-angle and loop integrals for these
potentials.  The approach used is similar to that of section
\ref{loopints}.  Analyzing the potentials reveals that each of the terms
can be written in the following schematic form,
\begin{eqnarray}
V_{\text{TBE:UX,i}}(k^+,k_\perp;p^+,p_\perp) &=&
\int_0^\infty \frac{q_\perp dq_\perp}{2(2\pi)^3} \int_0^1 dz \,
J(k^+,q^+,p^+)
I_i(k^+,k_\perp,q^+,q_\perp,p^+,p_\perp) \\
I_i(k^+,k_\perp,q^+,q_\perp,p^+,p_\perp) &=& 
\int_0^{2\pi} 
\frac{d\phi_q d\phi}{A_{1,i} + B_{1,i} \cos \phi_q + C_{1,i} \cos \phi}
\left( \frac{1}{A_2 + B_2 \cos \phi_q} \right)^2,
\end{eqnarray}
where $\phi=-\phi_q+\phi_p$, and the $A$'s, $B$'s, and $C$'s may have
dependence on $k^+$, $k_\perp$, $p^+$, $p_\perp$, $q^+=zE$, and
$q_\perp$; they are independent of the azimuthal angles.  These factors
can be easily determined for each potential by examining the forms of
the original equations.  The rotational invariance of the potential
about the three-axis allows the $\phi_k$ integration to be done
trivially.  The $\phi$ integral is easily done to obtain
\begin{eqnarray}
I(k^+,k_\perp,q^+,q_\perp,p^+,p_\perp) &=& -2\pi
\int_0^{2\pi} d\phi_q
\frac{1}{\sqrt{a_i^2-C_{1,i}^2}}
\left( \frac{1}{A_2 + B_2 \cos \phi_q} \right)^2,
\end{eqnarray}
where
\begin{eqnarray}
a_i &=& A_{1,i} + B_{1,i} \cos \phi_q.
\end{eqnarray}

Further simplification is possible for $V_{\text{TBE:UX2}}$, since for
that potential $B_{1,2}=0$,
\begin{eqnarray}
I(k^+,k_\perp,q^+,q_\perp,p^+,p_\perp) &=& \frac{(2\pi)^2 A_2}{
\sqrt{A_{1,2}^2-C_{1,2}^2}\sqrt{A_2^2-B_2^2}(A_2^2-B_2^2)}.
\end{eqnarray}
The techniques discussed in the previous section are used to do the
remaining loop integrals.

\begin{figure}[!ht]
\begin{center}
\epsfig{angle=0,width=5.0in,height=4.0in,file=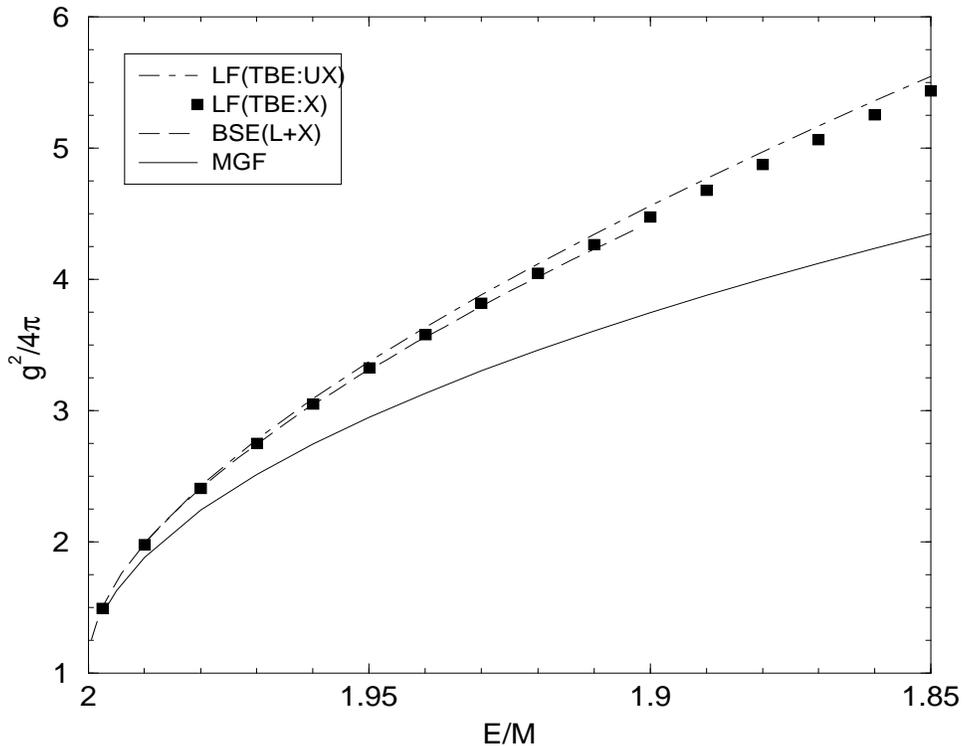}
\caption{The spectra are plotted for the potential derived from the TBE
truncation of the uncrossed approximation (OBE+TBE:SB+TBE:UX, denoted in
the figure by TBE:UX), the TBE potential (OBE+TBE, denoted by TBE:X),
the ladder plus crossed Bethe-Salpeter equation (BSE(L+X)), and the
modified-Green's-function potential (MGF).  Except for the MGF spectrum,
all the curves lie close to each other.
\label{uxtestfig}}
\end{center}
\end{figure}

The spectra for the OBE+TBE:SB+TBE:UX potential can be calculated and
compared to the OBE+TBE:SB+TBE:X potential (which is the same as the
OBE+TBE potential), the modified-Green's-function potential, and the
ladder plus crossed Bethe-Salpeter equation.  The spectra are plotted in
Fig.~\ref{uxtestfig}.  The spectra for the TBE:UX, TBE:X and BSE all lie
close to each other, which shows that that the uncrossed approximation
is valid.

This shows that the important approximation in the
modified-Green's-function approach is not the uncrossed approximation,
but the addition of the extra interaction terms in
Eq.~(\ref{ladderuncrossBSE}) which serve to mimic the higher order
interactions.


\begin{references}

\bibitem{Alexa:1999fe}
L.~C.~Alexa {\it et al.}  [Jefferson Lab Hall A Collaboration],
Phys.\ Rev.\ Lett.\  {\bf 82}, 1374 (1999)
[nucl-ex/9812002].

\bibitem{Dirac:1949cp}
P.~A.~Dirac,
Rev.\ Mod.\ Phys.\  {\bf 21}, 392 (1949).

\bibitem{Brodsky:1998de}
S.~J.~Brodsky, H.~Pauli and S.~S.~Pinsky,
Phys.\ Rept.\  {\bf 301}, 299 (1998)
[hep-ph/9705477].

\bibitem{Harindranath:1996hq}
A.~Harindranath,
hep-ph/9612244.

\bibitem{Heinzl:1998kz}
T.~Heinzl,
hep-th/9812190.

\bibitem{Miller:1997cr}
G.~A.~Miller,
Phys.\ Rev.\  {\bf C56}, 2789 (1997)
[nucl-th/9706028].

\bibitem{Cooke:1999yi}
J.~R.~Cooke, G.~A.~Miller and D.~R.~Phillips,
nucl-th/9910013.

\bibitem{Simonov:1993kp}
Y.~A.~Simonov and J.~A.~Tjon,
Annals Phys.\  {\bf 228}, 1 (1993).

\bibitem{Nambu:1950vt}
Y.~Nambu,
Prog.\ Theor.\ Phys.\ {\bf 5}, 614 (1950).
\bibitem{Schwinger:1951ex}
J.~Schwinger,
Proc.\ Nat.\ Acad.\ Sci.\ {\bf 37}, 452 (1951);
{\bf 37}, 455 (1951).
\bibitem{Gell-Mann:1951rw}
M.~Gell-Mann and F.~Low,
Phys.\ Rev.\  {\bf 84}, 350 (1951).
\bibitem{Salpeter:1951sz}
E.~E.~Salpeter and H.~A.~Bethe,
Phys.\ Rev.\  {\bf 84}, 1232 (1951).

\bibitem{Gross:1982nz}
F.~Gross,
Phys.\ Rev.\  {\bf C26}, 2203 (1982).

\bibitem{LevineWright1}
M.J.~Levine, J.A.~Tjon, and J.~Wright,
Phys.~Rev.~Lett.~{\bf 16}, 962 (1966),
\bibitem{LevineWright2}
M.J.~Levine, J.~Wright, and J.A.~Tjon,
Phys.~Rev.~{\bf 154}, 1433 (1967).
\bibitem{LevineWright3}
M.J.~Levine and J.~Wright,
Phys.~Rev.~{\bf D2}, 2509 (1970).

\bibitem{Carbonell:1998rj}
J.~Carbonell, B.~Desplanques, V.~A.~Karmanov and J.~F.~Mathiot,
Phys.\ Rept.\  {\bf 300}, 215 (1998)
[nucl-th/9804029].

\bibitem{Schoonderwoerd:1998pk}
N.~C.~Schoonderwoerd, B.~L.~Bakker and V.~A.~Karmanov,
Phys.\ Rev.\  {\bf C58}, 3093 (1998)
[hep-ph/9806365].

\bibitem{Baym}
G.~Baym,
Phys.\ Rev.\ {\bf 117}, 886 (1960).

\bibitem{Wivoda:1993qr}
J.~J.~Wivoda and J.~R.~Hiller,
Phys.\ Rev.\  {\bf D47}, 4647 (1993).

\bibitem{Savkli:1999rw}
C.~Savkli, J.~Tjon and F.~Gross,
Phys.\ Rev.\  {\bf C60}, 055210 (1999)
[hep-ph/9906211].

\bibitem{Rosenfelder:1996bd}
R.~Rosenfelder and A.~W.~Schreiber,
Phys.\ Rev.\  {\bf D53}, 3337 (1996)
[nucl-th/9504002];
{\bf 53}, 3354 (1996)
[nucl-th/9504005].

\bibitem{Sales:1999ec}
J.~H.~Sales, T.~Frederico, B.~V.~Carlson and P.~U.~Sauer,
nucl-th/9909029.

\bibitem{Ligterink:1995tm}
N.~E.~Ligterink and B.~L.~Bakker,
Phys.\ Rev.\  {\bf D52}, 5954 (1995)
[hep-ph/9412315].

\bibitem{Nieuwenhuis:1996mc}
T.~Nieuwenhuis and J.~A.~Tjon,
Phys.\ Rev.\ Lett.\  {\bf 77}, 814 (1996)
[hep-ph/9606403].

\bibitem{WickCutkosky}
G.~C.~Wick,
Phys.\ Rev.\  {\bf 96}, 1124 (1954);
R.~E.~Cutkosky,
{\it ibid.}
{\bf 96}, 1135 (1954).

\bibitem{Terentev:1976jk}
M.~V.~Terentev,
Sov.\ J.\ Nucl.\ Phys.\  {\bf 24}, 106 (1976).

\bibitem{klein}
A.~Klein, Phys.~Rev.~{\bf 90}, 1101 (1953).

\bibitem{Phillips:1996eb}
D.~R.~Phillips and S.~J.~Wallace,
Phys.\ Rev.\  {\bf C54}, 507 (1996)
[nucl-th/9603008].

\bibitem{Lahiff:1997bj}
A.~D.~Lahiff and I.~R.~Afnan,
Phys.\ Rev.\  {\bf C56}, 2387 (1997)
[nucl-th/9708037].

\bibitem{Chang:1969bh}
S.~Chang and S.~Ma,
Phys.\ Rev.\  {\bf 180}, 1506 (1969).

\bibitem{Krautgartner:1992xz}
M.~Krautgartner, H.~C.~Pauli and F.~Wolz,
Phys.\ Rev.\  {\bf D45}, 3755 (1992).

\bibitem{Trittmann:1997ga}
U.~Trittmann and H.~Pauli,
hep-th/9705021.

\bibitem{Eden:1996ey}
J.~A.~Eden and M.~F.~Gari,
Phys.~Rev.~{\bf C53}, 1510 (1996)
[nucl-th/9601025],
S.~Okubo, Prog.~Theor.~Phys.~{\bf 12}, 603 (1954).

\bibitem{Blankenbecler:1966gx}
R.~Blankenbecler and  R.~Sugar,
Phys.~Rev.~{\bf 142}, 1051 (1966).

\bibitem{Gross:1969rv}
F.~Gross,
Phys.~Rev.~{\bf 186}, 1448 (1969).

\bibitem{wallace} S.~J.~Wallace, in
{\it Nuclear and Particle Physics on the Light Cone (LAMPF Workshop), 1988},
edited by M.B.~Johnson and L.S.~Kisslinger
(World Scientific, Singapore, 1989), p.\ 477.

\bibitem{Wallace:1989nm}
S.~J.~Wallace and V.~B.~Mandelzweig,
Nucl.\ Phys.\  {\bf A503}, 673 (1989).

\bibitem{Woloshyn:1973} 
R.M.~Woloshyn and A.D.~Jackson,
Nucl.~Phys.~{\bf B64}, 269 (1973).

\bibitem{Phillips:1996ed}
D.~R.~Phillips and I.~R.~Afnan,
Phys.\ Rev.\  {\bf C54}, 1542 (1996)
[nucl-th/9605004].

\bibitem{schwartz}
C.~Schwartz,
Phys.~Rev.~{\bf 137}, B717 (1965).

\bibitem{Theussl:1999xq}
L.~Theussl and B.~Desplanques,
nucl-th/9908007.

\bibitem{dandata}
D.R.~Phillips (private communication).

\bibitem{press} W.H.~Press, S.A.~Teukolsky, W.T.~Vetterling, and
B.P.~Flannery,
{\it Numerical Recipes in C, second edition}
(Cambridge University Press, New York, 1992), p.~147.

\bibitem{eispack}
B.T.~Smith {\it et al},
Matrix Eigensystem Routines -- EISPACK
Guide, Lecture Notes in Computer Science, Vol.~6, Second Edition,
Springer-Verlag, New York, Heidelberg, Berlin, 1976;
B.S.~Garbow {\it et al},
Matrix Eigensystem Routines -- EISPACK Guide Extension, Lecture Notes in
Computer Science, Vol.~51, Springer-Verlag, New York, Heidelberg,
Berlin, 1977.

\end{references}
\end{document}